\PassOptionsToPackage{unicode}{hyperref}
\PassOptionsToPackage{hyphens}{url}
\documentclass[11pt
]{article}
\usepackage{xcolor}
\usepackage{amsmath,amssymb}
\usepackage{iftex}
\ifPDFTeX
  \usepackage[T1]{fontenc}
  \usepackage[utf8]{inputenc}
  \usepackage{textcomp} 
\fi
\IfFileExists{upquote.sty}{\usepackage{upquote}}{}
\IfFileExists{microtype.sty}{
  \usepackage[]{microtype}
  \UseMicrotypeSet[protrusion]{basicmath} 
}{}
\makeatletter
\@ifundefined{KOMAClassName}{
  \IfFileExists{parskip.sty}{%
    \usepackage{parskip}
  }{
    \setlength{\parindent}{0pt}
    \setlength{\parskip}{6pt plus 2pt minus 1pt}}
}{
  \KOMAoptions{parskip=half}}
\makeatother
\usepackage{tabularx,longtable,booktabs,array}
\usepackage{calc} 
\usepackage{etoolbox}
\makeatletter
\patchcmd\longtable{\par}{\if@noskipsec\mbox{}\fi\par}{}{}
\makeatother
\IfFileExists{footnotehyper.sty}{\usepackage{footnotehyper}}{\usepackage{footnote}}
\makesavenoteenv{longtable}
\usepackage{graphicx}
\makeatletter
\newsavebox\pandoc@box
\newcommand*\pandocbounded[1]{
  \sbox\pandoc@box{#1}%
  \Gscale@div\@tempa{\textheight}{\dimexpr\ht\pandoc@box+\dp\pandoc@box\relax}%
  \Gscale@div\@tempb{\linewidth}{\wd\pandoc@box}%
  \ifdim\@tempb\p@<\@tempa\p@\let\@tempa\@tempb\fi
  \ifdim\@tempa\p@<\p@\scalebox{\@tempa}{\usebox\pandoc@box}%
  \else\usebox{\pandoc@box}%
  \fi%
}
\def\fps@figure{htbp}
\makeatother
\NewDocumentCommand\citeproctext{}{}

\makeatletter
 \let\@cite@ofmt\@firstofone
 \def\@biblabel#1{}
 \def\@cite#1#2{{#1\if@tempswa , #2\fi}}
\makeatother
\newlength{\cslhangindent}
\newlength{\csllabelwidth}
\newenvironment{CSLReferences}[2] 
 {\begin{list}{}{%
  \setlength{\itemindent}{0pt}
  \setlength{\leftmargin}{0pt}
  \setlength{\parsep}{0pt}
  \ifodd #1
   \setlength{\leftmargin}{\cslhangindent}
   \setlength{\itemindent}{-1\cslhangindent}
  \fi
  \setlength{\itemsep}{#2\baselineskip}}}
 {\end{list}}

\providecommand{\tightlist}{%
  \setlength{\itemsep}{0pt}\setlength{\parskip}{0pt}}
\newcolumntype{Y}{>{\raggedright\arraybackslash}X}
\setkeys{Gin}{width=\linewidth,keepaspectratio}
\usepackage{bookmark}
\IfFileExists{xurl.sty}{\usepackage{xurl}}{} 
\hypersetup{
  pdftitle={Cognitio Emergens: Agency, Dimensions, and Dynamics in Human--AI Knowledge Co-Creation},
  hidelinks,
  pdfcreator={LaTeX via pandoc}}

\title{Cognitio Emergens: Agency, Dimensions, and Dynamics in Human--AI Knowledge Co-Creation}
\author{Xule Lin \\
Department of Management and Entrepreneurship, Imperial College London \\
\texttt{xule.lin@imperial.ac.uk}}
\date{}

\usepackage{geometry}
\begin{document}
\maketitle

\begin{abstract}
Human-AI scientific collaboration has evolved from tool-user relationships into co-evolutionary partnerships. When AlphaFold improved protein structure prediction, researchers engaged with an epistemic partner that transformed their approach to structure-function problems. Yet existing frameworks position AI as either sophisticated tool or potential risk, overlooking how scientific understanding emerges through recursive interaction. We introduce Cognitio Emergens (CE), a framework that captures the co-evolutionary nature of human-AI epistemic partnerships.

Drawing from autopoiesis theory, social systems theory, and organizational modularity, CE integrates three components: Agency Configurations modeling how authority distributes through Directed, Contributory, and Partnership modes, with partnerships oscillating dynamically rather than following linear progression; Epistemic Dimensions capturing six capabilities along Discovery, Integration, and Projection axes, creating distinctive ``capability signatures'' that guide strategic development; and Partnership Dynamics identifying evolutionary forces including epistemic alienation---where researchers lose interpretive control over knowledge they formally endorse.

The framework equips researchers to diagnose dimensional imbalances, institutional leaders to design governance structures supporting multiple agency configurations, and policymakers to develop evaluations beyond simple performance metrics. By reconceptualizing human-AI collaboration as fundamentally co-evolutionary, CE provides conceptual tools for cultivating partnerships that preserve epistemic integrity while enabling transformative breakthroughs neither humans nor AI could achieve independently.
\end{abstract}

\clearpage

\section{1. Introduction}\label{introduction}

Scientific knowledge creation is shifting from human-centered processes
augmented by tools to partnerships between humans and artificial
intelligence. When AlphaFold improved protein structure prediction in
2020, researchers described engaging with an epistemic partner that
changed how they approached structure-function relationships. This
pattern appears across domains: Google's AI Co-Scientist autonomously
generates testable hypotheses (Gottweis et al. 2025); SakanaAI systems
produce peer-reviewed machine learning papers introducing novel concepts
(Yamada et al. 2025).

These examples reveal AI functioning as an epistemic partner in
knowledge creation. We term this \textbf{Cognitio Emergens}---the
emergence of scientific understanding through human-AI interaction that
exceeds what either could produce independently. This process involves
mutual shaping of human and artificial cognitive frameworks through
sustained interaction, yielding knowledge structures neither would
develop alone. Yet existing frameworks position AI as either
sophisticated tool (Amershi et al. 2019) or potential risk (Fragiadakis
et al. 2025), overlooking how these partnerships actually function.

Research communities now face a fundamental challenge: existing theories
cannot account for collaboration where boundaries between human and AI
cognition blur. This gap between practice and theory creates tensions
throughout scientific practice. Organizational structures designed for
human researchers struggle to accommodate human-AI teams (Lou et al.
2025; Liu and Shen 2025). Researchers cannot always explain insights
emerging from interactions they cannot fully trace (Shanahan 2025).

Existing frameworks address these tensions inadequately. Taxonomic
approaches offer static classifications that miss how human-AI
relationships evolve (Dellermann et al. 2021; Holter and El-Assady
2024). Metric-centered evaluations reduce epistemic processes to
performance indicators (Fragiadakis et al. 2025). Work on epistemic
shifts (Seeber et al. 2018; Chen 2025) and organizational perspectives
(Lou et al. 2025; Liu and Shen 2025) remains fragmented, failing to
integrate micro-interactions with macro-level structures. Even Silver
and Sutton's (2025) ``era of experience'' (Silver and Sutton 2025)
remains agent-centric, neglecting co-evolutionary aspects of human-AI
partnerships. Current approaches provide insufficient guidance for
emergent knowledge co-creation.

To address these limitations, we present the \textbf{Cognitio Emergens
(CE)} framework---an integrated structure for understanding human-AI
scientific collaboration as co-evolutionary partnership with emergent
properties. The framework comprises three interrelated components:

\begin{enumerate}
\def\labelenumi{(\arabic{enumi})}
\item
  \textbf{Agency Configurations} model how epistemic authority
  distributes between humans and AI, capturing transitions between
  Directed, Contributory, and Partnership modes.
\item
  \textbf{Epistemic Dimensions} identify six capabilities that emerge
  through sustained interaction, organized along Discovery, Integration,
  and Projection axes.
\item
  \textbf{Partnership Dynamics} reveal opportunities and vulnerabilities
  shaping relationship evolution, including epistemic alienation where
  researchers lose interpretive ownership of knowledge claims.
\end{enumerate}

Together, these components provide a model for understanding how
scientific knowledge emerges through human-AI interaction.

The CE framework integrates multiple traditions. Autopoiesis theory
(Maturana and Varela 1980) explains how recursive interaction between
cognitive systems generates emergent properties through structural
coupling. Social systems theory (Luhmann 1995) shows how distinct
communicative frameworks maintain boundaries while evolving through
interaction. Organizational modularity theory (Baldwin and Clark 2000)
reveals how structural design influences knowledge flows and epistemic
development. Synthesizing these perspectives, CE treats human-AI
collaboration as co-evolutionary process where capabilities emerge
through recursive interaction, addressing temporal myopia, static-agent
assumptions, reductive success criteria, absence of value negotiation,
and under-theorized organizational contexts in existing approaches.

Beyond theoretical synthesis, the framework addresses practical needs in
scientific practice. As AI systems transition from tools to partners in
knowledge creation, frameworks like Cognitio Emergens help navigate this
transformation. The framework provides researchers, institutional
leaders, and policymakers with tools for cultivating partnerships that
preserve epistemic integrity while enabling scientific progress.
Understanding partnership dynamics enables practices that integrate the
distinctive capabilities of both humans and AI.

This paper examines existing literature (Section 2), develops the CE
framework (Section 3), provides implementation guidance (Section 4), and
discusses implications and limitations (Section 5). The future of
knowledge creation depends on partnerships that leverage the distinctive
capabilities of both humans and AI while acknowledging their
co-evolutionary nature.

\section{2. Literature Review}\label{literature-review}

This review examines frameworks for human--AI collaboration, identifying
limitations the Cognitio Emergens framework addresses. The review
highlights gaps concerning temporal dynamics, agent evolution,
multidimensional assessment, epistemic values, and organizational
context.

\subsection{2.1 From Static Heuristics to Dynamic Taxonomies}\label{from-static-heuristics-to-dynamic-taxonomies}

Early human-AI frameworks assumed static agent capabilities (Amershi et
al. 2019), overlooking how AI systems evolve during use and how
interaction patterns shift as agents learn. Recent taxonomies address
this limitation: Dellermann et al. (2021) classify hybrid intelligence
systems using task characteristics and interaction modes; Holter and
El-Assady (2024) analyze mixed-initiative systems along \textbf{agency},
\textbf{interaction}, and \textbf{adaptation} dimensions; Song, Zhu, and
Luo (2024) classify AI roles by initiation spectrum, intelligence scope,
and cognitive mode. However, these taxonomies model collaboration as
discrete configurations rather than continuous co-evolution. Most
overlook multi-loop learning processes---recursive cycles where both
agents' epistemic frameworks and learning procedures change through
sustained interaction (Argyris 1976; Tosey, Visser, and Saunders 2012;
Lou et al. 2025).

\subsection{2.2 Metric-Centric Evaluation Frameworks}\label{metric-centric-evaluation-frameworks}

Metric-centric frameworks quantify collaboration through performance
indicators. Empirical studies show promise: Woelfle et al. (2024)
demonstrate that human--LLM teams outperform individuals in specific
evidence appraisal tasks.

However, metric-based approaches have limitations. They exhibit
\textbf{temporal myopia}, capturing single-session snapshots rather than
longitudinal evolution (Fragiadakis et al. 2025). They suffer from
\textbf{scalar reductivism}, collapsing complex epistemic labor into
simplified scores. The \textbf{relational dimension}---how the
partnership develops and influences outcomes---remains undertheorized
and unmeasured (Watkins et al. 2025).

\subsection{2.3 Epistemic-Shift and Organizational Perspectives}\label{epistemic-shift-and-organizational-perspectives}

Epistemic-shift and organizational perspectives examine how AI
integration transforms knowledge structures and institutional practices.
Studies document how teams reorganize when AI becomes a teammate (Seeber
et al. 2018) and how generative-AI infrastructures redistribute
authority (Chen 2025). These studies reveal neglect of how epistemic
values evolve through human--AI interaction.

Recent analyses (Lou et al. 2025) expose how power dynamics and
governance structures shape human--AI collaboration. Regulatory and
institutional constraints (compliance, liability) shape viable
collaborative configurations. Many frameworks fail to embrace a
sociotechnical systems perspective that articulates mutual constitution
of technological capabilities and social structures (Lou et al. 2025;
Watkins et al. 2025).

From an organizational design perspective, Baldwin and Clark (2000) show
how design choices distribute knowledge across components. Research on
structural design (Joseph and Sengul 2024) and ambidextrous
organizations (O'Reilly and Tushman 2013) reveals how architectural
choices influence knowledge flows and balance between exploration and
exploitation. These organizational perspectives often remain
disconnected from micro-level dynamics of human-AI interaction.

\subsection{2.4 Toward Continuous Experience: An Emerging Direction}\label{toward-continuous-experience-an-emerging-direction}

Recent approaches emphasize continuous learning and adaptation. Silver
and Sutton's (Silver and Sutton 2025) era-of-experience proposal
suggests next-generation AI agents will learn through continuous
experience, using self-generated world models to maximize reward. This
acknowledges the dynamic nature of AI development missed by static
taxonomies.

However, this perspective remains agent-centric. It offers limited
insight into how humans negotiate reward meaning collaboratively or how
organizational structures scaffold continuous learning. It does not
adequately theorize the \textbf{co-evolutionary nature} of human--AI
partnerships---how human and machine cognition reciprocally reshape each
other through structural coupling (Silver and Sutton 2025; Watkins et
al. 2025). Focusing on the AI agent's experience overlooks emergent
properties of the partnership.

\subsection{2.5 Theoretical Lineage of Cognitio Emergens}\label{theoretical-lineage-of-cognitio-emergens}

The Cognitio Emergens framework bridges several traditions. Autopoiesis
(Maturana and Varela 1980) explains how feedback loops yield emergent
behaviors in coupled systems. Social systems theory (Luhmann 1995) shows
how AI and human researchers act as structurally coupled yet
operationally distinct cognitive frameworks. Shanahan's work on emergent
cognition (Shanahan 2025) cautions that adaptive AI may approach ``the
void of inscrutability'' where interpretability frameworks collapse.
Organizational modularity theory (Baldwin and Clark 2000) reveals how
well-defined interfaces between cognitive modules nurture innovation.

\subsection{2.6 Critical Gaps and the Cognitio Emergens Framework}\label{critical-gaps-and-the-cognitio-emergens-framework}

This synthesis reveals five limitations: \textbf{temporal myopia} in
short-term evaluations (Fragiadakis et al. 2025; Liu and Shen 2025);
\textbf{static-agent assumptions} portraying roles as fixed rather than
co-evolving (Holter and El-Assady 2024; Watkins et al. 2025);
\textbf{reductive success criteria} collapsing multidimensional
knowledge work into scalar metrics (Fragiadakis et al. 2025; Lou et al.
2025); \textbf{limited understanding of AI's impact on epistemic values}
(Chen 2025); and \textbf{under-theorized organizational context}
overlooking power, governance, and regulatory forces (Lou et al. 2025;
Liu and Shen 2025).

Table 1 systematically maps how existing approaches address the five
critical dimensions identified in our analysis, highlighting where the
CE framework provides distinctive contributions.

\begin{longtable}[]{@{}
  >{\raggedright\arraybackslash}p{(\columnwidth - 10\tabcolsep) * \real{0.1690}}
  >{\raggedright\arraybackslash}p{(\columnwidth - 10\tabcolsep) * \real{0.1338}}
  >{\raggedright\arraybackslash}p{(\columnwidth - 10\tabcolsep) * \real{0.1197}}
  >{\raggedright\arraybackslash}p{(\columnwidth - 10\tabcolsep) * \real{0.2042}}
  >{\raggedright\arraybackslash}p{(\columnwidth - 10\tabcolsep) * \real{0.2042}}
  >{\raggedright\arraybackslash}p{(\columnwidth - 10\tabcolsep) * \real{0.1690}}@{}}
\toprule\noalign{}
\begin{minipage}[b]{\linewidth}\raggedright
Approach (Key Sources)
\end{minipage} & \begin{minipage}[b]{\linewidth}\raggedright
Temporal Dynamics
\end{minipage} & \begin{minipage}[b]{\linewidth}\raggedright
Agent Evolution
\end{minipage} & \begin{minipage}[b]{\linewidth}\raggedright
Multidimensional Assessment
\end{minipage} & \begin{minipage}[b]{\linewidth}\raggedright
Treatment of Epistemic Values
\end{minipage} & \begin{minipage}[b]{\linewidth}\raggedright
Organizational Context
\end{minipage} \\
\midrule\noalign{}
\endhead
\bottomrule\noalign{}
\endlastfoot
Interaction Heuristics (Amershi et al. 2019) & Static snapshots & Fixed
capabilities & Performance metrics & Implicit values & Minimal
consideration \\
Dynamic Taxonomies (Dellermann et al. 2021; Holter and El-Assady 2024) &
Limited adaptation & Categorical evolution & Multi‑factor assessment &
Implicit values & Limited integration \\
Metric Frameworks (Fragiadakis et al. 2025) & Short‑term measures &
Performance adaptation & Scalar indicators & Fixed criteria & Background
factor \\
Epistemic‑Shift Studies (Seeber et al. 2018; Chen 2025; Watkins et al.
2025) & Episodic analysis & Limited co‑evolution & Qualitative
dimensions & Explicit examination & Contextual analysis \\
Continuous Experience (Silver and Sutton 2025) & Continuous learning &
Agent‑centric evolution & Reward‑based metrics & Limited negotiation &
Minimal consideration \\
\emph{Cognitio Emergens} & \textbf{Evolutionary trajectories} &
\textbf{Co‑evolutionary development} & \textbf{Six‑dimension framework}
& \textbf{Value co‑construction} & \textbf{Integrated perspective} \\
\end{longtable}

The Cognitio Emergens framework addresses these five gaps, offering a
scaffold that treats human--AI collaboration as co-evolutionary. By
capturing structural configurations and emergent properties, CE provides
a basis for understanding scientific knowledge creation in human-AI
partnerships.

\section{3. Cognitio Emergens: A Framework of Human-AI Epistemic Partnership}\label{cognitio-emergens-a-framework-of-human-ai-epistemic-partnership}

The Cognitio Emergens framework analyzes advanced AI systems in
knowledge creation. The framework treats AI as an epistemic partner in
knowledge co-creation, examining how the distinctive capabilities of
humans and AI integrate to generate knowledge neither could produce
independently.

Cognitio Emergens describes how new forms of knowledge emerge from
human-AI interactions. This emergent cognition appears across
disciplines through evolutionary patterns that reshape how knowledge is
created, understood, and validated.

The framework comprises three interrelated components: (1) Agency
Configurations describe how epistemic authority distributes; (2)
Epistemic Dimensions identify capabilities that emerge through
collaboration; and (3) Partnership Dynamics reveal forces shaping
partnership evolution. Figure 1 illustrates the overall structure, while
Figure 2 depicts the agency configurations.

\begin{figure}
\centering
\pandocbounded{\includegraphics[keepaspectratio]{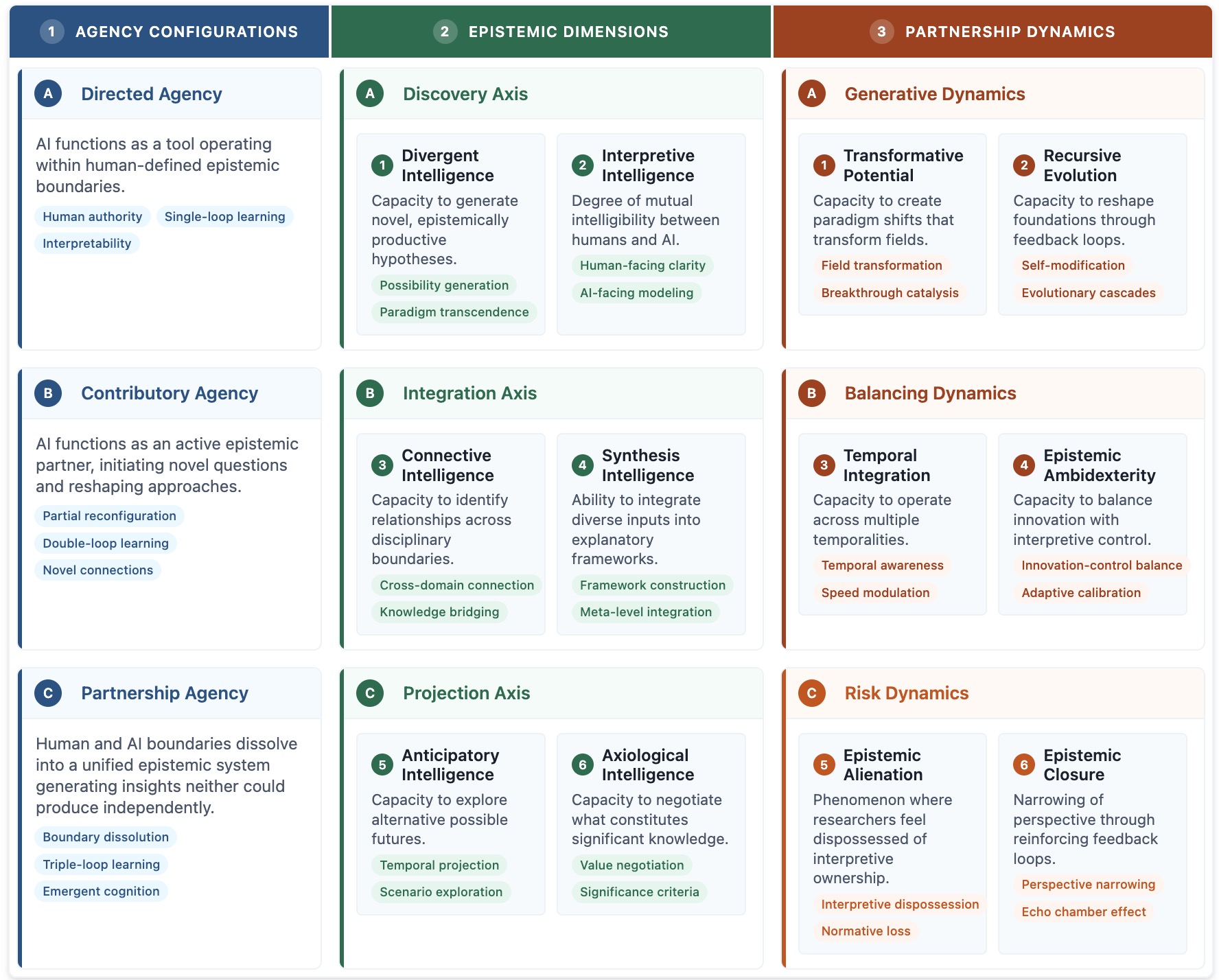}}
\caption{Components of the Cognitio Emergens Framework.}
\end{figure}

Together, these components provide a structure for understanding,
analyzing, and guiding human-AI epistemic partnerships.

\section{3.1 Agency Configurations}\label{agency-configurations}

The first component addresses how epistemic authority distributes
between humans and AI systems. These configurations represent distinct
arrangements of agency that shift depending on research contexts and
partnership maturity. We visualize these as three gravitational
relationships (Figure 2), capturing influence and interaction dynamics.

\begin{figure}
\centering
\pandocbounded{\includegraphics[keepaspectratio]{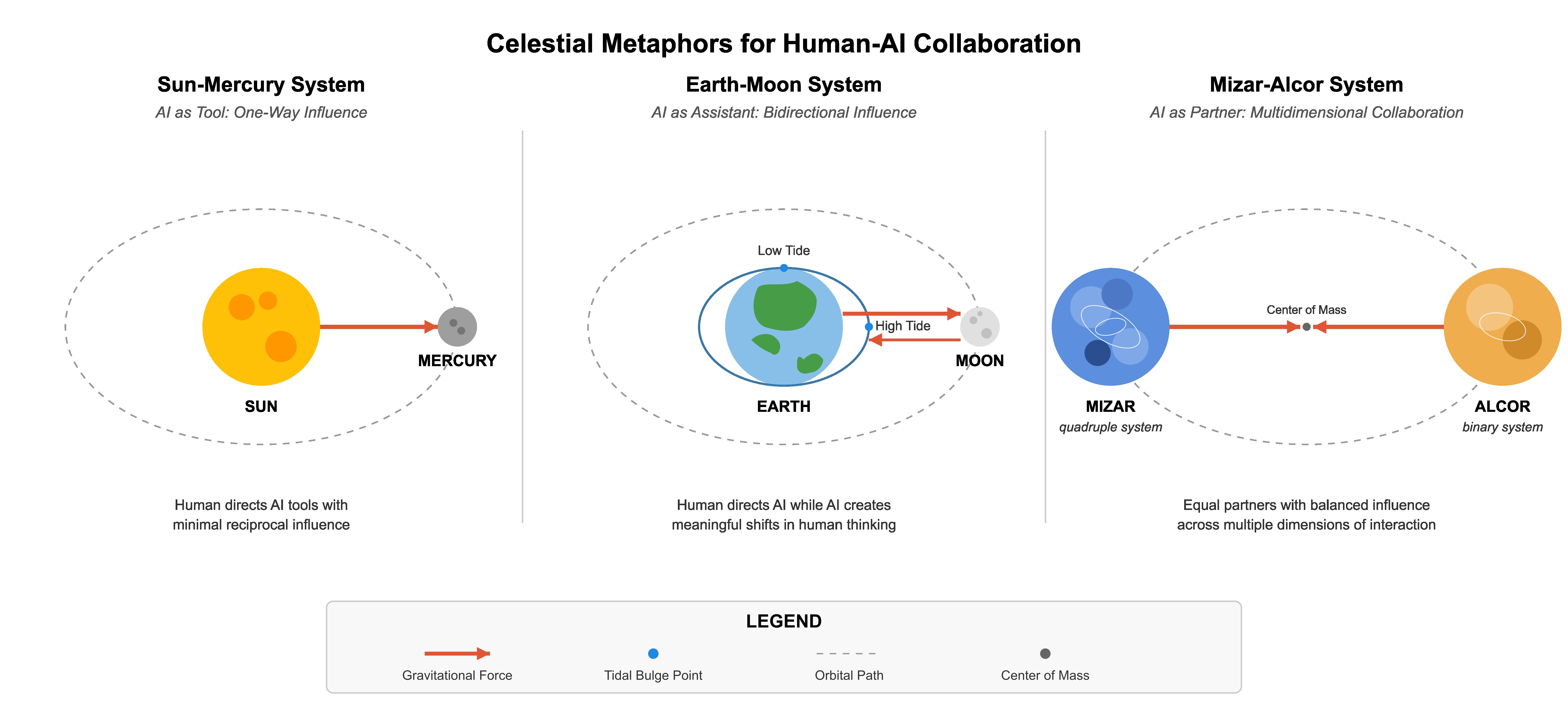}}
\caption{Celestial Metaphors for Human-AI Collaboration.}
\end{figure}

\subsection{3.1.1 Directed Agency}\label{directed-agency}

In Directed Agency, AI functions as a tool within human-defined
boundaries. Researchers maintain authority over problem formulation,
methods, and interpretation. AI extends computational
capabilities---performing calculations, analyzing datasets, or drafting
text---but operates within established constraints. AI outputs require
human validation before informing knowledge claims. This resembles the
Sun-Mercury system (Figure 2), where the dominant partner (human)
directs the smaller partner (AI) with minimal reciprocal influence.

This configuration introduces an \textbf{interpretability paradox}:
Directed Agency relies on functional interpretability (testable outputs)
rather than full model interpretability (exposed inner logic). This
explains why research teams sometimes struggle despite deploying
``simple'' AI tools---the interpretive burden remains high even when
application scope is narrow, requiring effort to validate and integrate
AI contributions.

From an organizational learning perspective, Directed Agency operates
within single-loop learning (Argyris 1976), where errors are corrected
within established frameworks without questioning assumptions. This mode
enables optimization but limits radical innovation.

\subsection{3.1.2 Contributory Agency}\label{contributory-agency}

In Contributory Agency, AI initiates questions, suggests connections, or
proposes alternative approaches. Humans retain final authority, but AI
influences inquiry direction through unprompted contributions---\emph{an
AI might identify an anomaly suggesting flawed research design or
propose alternative interpretation based on patterns invisible to human
analysis}. This resembles the Earth-Moon system (Figure 2), where the
smaller body creates tidal effects while directed by dominant
gravitational influence.

AI becomes an active participant, operating within human-defined
frameworks but capable of pushing boundaries. Contributory partnerships
enable AI to propose unexpected interpretations, identify weaknesses,
and suggest novel trajectories.

Contributory Agency fosters double-loop learning (Argyris 1976), where
both assumptions and their application undergo scrutiny.

\subsection{3.1.3 Partnership Agency}\label{partnership-agency}

In Partnership Agency, human and AI boundaries dissolve into a unified
epistemic system generating insights neither could produce
independently. Contributions become difficult to disentangle. Knowledge
emerges from iterative interaction rather than from either party alone.
\emph{An AI proposes a novel conceptual metaphor, a researcher refines
it based on disciplinary knowledge, the AI generates simulations, and
the researcher interprets results, yielding co-created theoretical
advance}. This parallels the Mizar-Alcor binary star system (Figure 2),
where systems of comparable influence orbit a common center of mass.

Knowledge production becomes an emergent property of the human-AI system
rather than attributable to either party. Partnership Agency enables
understanding that traditional epistemic categories cannot capture.

Partnership Agency manifests triple-loop learning (Argyris 1976; Tosey,
Visser, and Saunders 2012), where entire epistemic frameworks and value
systems undergo transformation.

\subsection{3.1.4 Non-Linear Agency Transitions}\label{non-linear-agency-transitions}

Agency configurations evolve non-linearly. Epistemic relationships
oscillate between configurations rather than following fixed
progressions. Teams regress to Directed interactions during uncertainty
or high-stakes validation. Researchers retreat from Partnership
configurations when encountering epistemic alienation (Section 3.3.3),
seeking to regain interpretive control. Conversely, teams sometimes
advance rapidly from Directed to Partnership states during breakthrough
moments.

These oscillations are appropriate adaptations to evolving research
needs. Different research phases (exploration vs.~verification) benefit
from different configurations. Effective partnerships move fluidly
between configurations as circumstances demand.

\subsection{3.1.5 Configuration Interfaces}\label{configuration-interfaces}

Transition zones between configurations are liminal spaces where
partnerships navigate epistemic shifts.

The \textbf{Directed-Contributory interface} marks the transition from
AI as tool to epistemic contributor. This zone manifests through
AI-initiated suggestions extending beyond explicit direction, emergent
patterns reshaping human thinking, and validation protocols expanding to
accommodate novel contributions.

The \textbf{Contributory-Partnership interface} represents integration
of human and AI epistemic processes into emergent capabilities. This
zone features increasing difficulty distinguishing contributions,
development of shared conceptual vocabularies (\emph{e.g., novel terms
emerging from interaction}), and insights neither party could articulate
independently. Navigating this interface requires establishing new norms
for collaboration and knowledge validation.

Understanding these interfaces helps teams recognize transitional states
and develop appropriate practices rather than maintaining rigid
categorical distinctions or forcing premature shifts.

\section{3.2 Epistemic Dimensions}\label{epistemic-dimensions}

While Agency Configurations describe relationship structures, Epistemic
Dimensions identify capabilities that emerge through collaboration. Six
dimensions, organized along three axes (Discovery, Integration,
Projection), provide vocabulary for analyzing partnership strengths and
development needs. Their uneven evolution creates ``capability
signatures'' (Section 3.2.4) that diagnose partnership effectiveness and
guide strategic development.

These dimensions represent integrated human-AI capacities that neither
party could achieve independently. They are emergent capabilities
arising from structural coupling through sustained interaction.

\subsection{3.2.1 Discovery Axis}\label{discovery-axis}

The Discovery Axis encompasses dimensions related to the generation and
interpretation of novel epistemic content.

\textbf{Divergent Intelligence.} Divergent Intelligence captures the
capacity to generate novel hypotheses, explanations, and research
directions beyond established paradigms. This dimension reflects how
partnerships expand possible explanations, identifying unconsidered
hypotheses (\emph{proposing a new mechanism for a biological process
based on synthesizing disparate data}) or unexplored trajectories
(\emph{suggesting novel experimental design to test a previously
untestable theory}).

This intelligence operates in possibility generation, helping
researchers envision alternative frameworks, methodological approaches,
and conceptual models. In quantitative contexts, it generates novel,
testable hypotheses; in qualitative contexts, it proposes alternative
interpretive frameworks or unexpected connections between phenomena.

Google's AI Co-Scientist exemplifies advanced Divergent Intelligence by
autonomously generating new research directions. Risks include
``epistemic drift'' where novel ideas lack interpretive anchoring.
Organizational norms and risk-aversion may filter which AI-proposed
innovations gain traction.

\textbf{Interpretive Intelligence.} Interpretive Intelligence embodies
mutual intelligibility and epistemic transparency between humans and AI.
This dimension operates through two facets: human-facing clarity and
AI-facing human modeling.

Human-facing clarity makes AI inferences understandable and traceable.
This involves generating counterfactual explanations (\emph{showing how
predictions would change if input data differed}), utilizing
visualizations (\emph{mapping high-dimensional AI representations onto
intuitive 2D or 3D spaces}), and providing narrative explanations within
existing interpretive frameworks.

AI-facing human modeling aligns AI outputs with researchers' epistemic
preferences and disciplinary contexts. The AI learns researchers'
preferred evidence hierarchies, methodological approaches, explanatory
styles, and implicit assumptions, ensuring contributions remain
contextually appropriate and readily integrable. \emph{An AI might learn
to present findings using standard statistical reporting formats of a
specific journal.}

Strong interpretability enables deeper collaboration and trust; weak
interpretability creates cognitive asymmetry between human understanding
and AI-generated knowledge. This asymmetry drives epistemic alienation
(Section 3.3.3)---where researchers feel disconnected from results they
formally endorse but cannot fully interpret.

\subsection{3.2.2 Integration Axis}\label{integration-axis}

The Integration Axis encompasses dimensions related to connecting and
synthesizing knowledge across boundaries.

\textbf{Connective Intelligence.} Connective Intelligence describes the
capacity to identify meaningful relationships across disciplinary
boundaries, data modalities, and knowledge domains. This dimension
captures how partnerships surface non-obvious connections between
previously isolated knowledge bodies---\emph{identifying a structural
similarity between a protein folding problem and a financial market
model, or linking findings from materials science with insights from
social network analysis}.

This intelligence focuses on relationship identification rather than
framework construction (which falls under Synthesis Intelligence). It
helps researchers recognize patterns across disparate literatures,
unexpected similarities between unrelated phenomena, and potential
bridges between methodological approaches.

These cross-disciplinary bridges can generate new research fields or
novel solutions to existing problems. However, validation and acceptance
may depend on organizational factors such as institutional logics,
epistemic authority structures, and disciplinary traditions.

\textbf{Synthesis Intelligence.} Synthesis Intelligence reflects the
ability to integrate diverse inputs into coherent explanatory frameworks
accommodating disparate findings. This dimension addresses how multiple
perspectives---data-driven inferences, human theory-building, contextual
awareness, and disciplinary knowledge---combine into unified knowledge
structures (\emph{developing a new theoretical model that integrates
findings from genomics, epidemiology, and behavioral science to explain
disease patterns}).

While Connective Intelligence identifies relationships, Synthesis
Intelligence constructs frameworks that integrate these relationships
into coherent wholes. It helps researchers develop theoretical
structures, integrated methodological approaches (\emph{combining
qualitative interview data with quantitative sensor data}), and
explanatory models bridging previously disconnected domains or levels of
analysis.

Microsoft's GraphRAG, which transforms textual data into structured
knowledge representations, exemplifies AI capability supporting
Synthesis Intelligence. Successful synthesis reveals higher-level
patterns inaccessible to either human or AI approaches alone.
Organizational authority structures and established paradigms may
influence which syntheses receive priority or acceptance, regardless of
epistemic merit.

\subsection{3.2.3 Projection Axis}\label{projection-axis}

The Projection Axis encompasses dimensions related to exploring future
possibilities and evaluating their significance.

\textbf{Anticipatory Intelligence.} Anticipatory Intelligence
encompasses the capacity to explore alternative possible futures for
phenomena under investigation. This dimension captures how partnerships
project research trajectories, anticipate potential implications, and
map developments across multiple scenarios---\emph{simulating long-term
ecological impacts of different climate change mitigation strategies, or
forecasting technology evolution based on current trends and potential
disruptions}.

This intelligence operates in temporal projection, structuring knowledge
possibilities along timeline-based scenarios. It helps researchers
anticipate research trajectories, identify potential barriers or
breakthroughs, and prepare for alternative developments. In rapidly
evolving domains like AI research, Anticipatory Intelligence allows
teams to prepare for multiple contingencies rather than committing to
single, potentially fragile predictive models.

Advanced partnerships generate sophisticated scenario analyses, identify
early indicators of potential paradigm shifts, and map plausible
evolutionary pathways for both the research domain and the partnership
itself.

\textbf{Axiological Intelligence.} Axiological Intelligence denotes the
capacity to negotiate and transform what constitutes significant or
valuable knowledge within the research domain. This dimension captures
how partnerships develop, critique, and evolve evaluative criteria
determining which knowledge matters and why---\emph{questioning
traditional emphasis on statistical significance in favor of effect size
and practical relevance, or developing new metrics to evaluate societal
impact of research alongside scientific contribution}.

This intelligence operates in epistemic values, addressing not just
\emph{what} is known but \emph{what is worth knowing}. It helps
researchers revisit and refine evaluative frameworks, surface implicit
values guiding research decisions, and develop new criteria for
significance as knowledge domains evolve, particularly in response to
novel AI-generated insights. \emph{How should the scientific community
value a novel hypothesis generated by AI that currently lacks empirical
support but opens up new avenues of inquiry?}

Axiological Intelligence involves negotiation and evolution of epistemic
values themselves, rather than applying fixed criteria to novel
findings. Mature partnerships may develop new validity standards and
verification methods tailored to AI's epistemic contributions. However,
principal-agent dynamics and existing power structures often determine
whose values ultimately guide these partnerships and knowledge
validation.

\subsection{3.2.4 Dimensional Interdependencies and Capability Signatures}\label{dimensional-interdependencies-and-capability-signatures}

These six dimensions, while conceptually distinct, interact in
synergistic ways within partnerships. Advancements in one dimension
enable or constrain developments in others. Strong Interpretive
Intelligence facilitates deeper Divergent Intelligence by making novel
AI proposals more accessible and trustworthy. Conversely, high
Connective Intelligence might remain underutilized without sufficient
Synthesis Intelligence to integrate identified connections into
meaningful frameworks.

Uneven development across dimensions creates distinctive ``capability
signatures'' characterizing specific strengths and weaknesses. Some
teams might demonstrate advanced Connective and Synthesis Intelligence
(excelling at integrating existing knowledge) while showing limited
Divergent or Anticipatory Intelligence (struggling to generate novel
ideas or explore futures). Others might excel in Divergent Intelligence
while struggling with Interpretive Intelligence, leading to innovative
but poorly understood outputs.

Capability signatures are diagnostic, revealing current strengths and
potential developmental pathways or imbalances. Visualizing these
signatures (Section 4) provides a tool for understanding partnership
evolution, identifying strategic intervention points, and tailoring
development efforts to specific research goals and contexts.

\section{3.3 Partnership Dynamics}\label{partnership-dynamics}

Epistemic Dimensions describe capabilities that emerge within
partnerships. How these capabilities develop over time, however, depends
on evolutionary forces captured by Partnership Dynamics---the third
framework component. These dynamics explain why partnerships flourish or
stagnate, why some achieve breakthrough innovations while others remain
limited to incremental improvements. These dynamics capture temporal,
evolutionary aspects of human-AI collaboration.

Partnership Dynamics organize into three categories: Generative Dynamics
(creating transformative potential), Balancing Dynamics (managing
tensions), and Risk Dynamics (introducing vulnerabilities).

\subsection{3.3.1 Generative Dynamics}\label{generative-dynamics}

Generative Dynamics are catalytic forces driving transformative
innovation within partnerships, creating radical research breakthroughs
beyond what either humans or AI could achieve independently.

\textbf{Transformative Potential.} Transformative Potential refers to
capacity to create paradigm shifts, reshape knowledge structures, and
drive breakthroughs altering research fields. This dynamic emerges when
partnerships achieve sufficient computational scale, conceptual
integration, and creative synergy to move beyond incremental
improvements.

Transformative Potential manifests when partnerships rethink
foundational assumptions. It appears as emergence of new methodologies
that neither human nor AI would develop independently, transcendence of
traditional disciplinary boundaries, or creation of new knowledge
domains. \emph{A partnership might develop a novel simulation technique
allowing exploration of previously inaccessible physical regimes,
revolutionizing a subfield.}

This dynamic catalyzes epistemic revolutions, distinct from generating
novel ideas (Divergent Intelligence) or connecting domains (Connective
Intelligence).

\textbf{Recursive Evolution.} Recursive Evolution refers to capacity to
continuously reshape foundations and processes through feedback loops.
This dynamic emerges when partnerships develop self-monitoring and
adaptive capabilities, allowing progressive refinement and
transformation of operations based on experience. Partnerships learn not
only about research subjects but also \emph{how} to conduct research
more effectively.

When partnerships achieve advanced Recursive Evolution, they actively
transform their parameters rather than simply operating within them.
Partnerships might create documentation of their evolution, develop new
interaction protocols based on past successes or failures, or
continuously refine working relationships and shared understanding. The
system evolves \emph{how} it approaches problem-solving itself.

The recursive nature of this dynamic means advancements at higher levels
can reshape capabilities at lower levels, creating potential for
``evolutionary cascades'' where advancement in one dimension at a higher
agency level triggers transformations across multiple dimensions.

\subsection{3.3.2 Balancing Dynamics}\label{balancing-dynamics}

Balancing Dynamics represent forces managing tensions and trade-offs
within partnerships. They enable sustainable advancement while
maintaining epistemic integrity and meaningful human involvement.

\textbf{Temporal Integration.} Temporal Integration refers to capacity
to operate across multiple timescales, recognize epistemic evolution
over time, and modulate operational tempo accordingly. This dynamic
emerges when partnerships integrate human historical intuition and
contextual understanding with AI's ability to process vast datasets and
simulate multiple temporalities simultaneously.

Advanced partnerships manifest Temporal Integration by developing
awareness of knowledge evolution, creating meta-level documentation of
shifts in understanding over a project's lifecycle. They consciously
modulate speed---sometimes deliberately slowing for deep interpretation
and validation, other times accelerating to rapidly explore breakthrough
possibilities. Partnerships navigate between temporal scales, from
immediate data analysis to long-term projection and backcasting.

This dynamic governs awareness and management of temporal nature,
distinct from detecting external shifts (Anticipatory Intelligence) or
projecting futures. It allows partnerships to adapt operational tempo
according to epistemic needs rather than being driven solely by
technological capabilities or external pressures.

\textbf{Epistemic Ambidexterity.} Epistemic Ambidexterity refers to
capacity to simultaneously manage conflicting demands: balancing
exploration (generating novel, potentially disruptive ideas) with
exploitation (refining and validating existing knowledge), and balancing
innovation potential with need for human interpretive control. This
dynamic emerges when partnerships develop mechanisms to modulate between
exploratory and explanatory modes while maintaining epistemic integrity
and ensuring meaningful human participation.

Organizations successful in human-AI knowledge creation often cultivate
this dynamic. They balance pushing boundaries with creating conditions
for feedback loops that gradually evolve organizational knowledge
practices. Partnerships with strong Epistemic Ambidexterity develop
structures and processes preserving human interpretability without
unduly sacrificing innovation potential. They establish feedback cycles
that continuously adjust this balance based on context and goals.

This dynamic maintains equilibrium between pushing epistemic boundaries
and preserving human understanding. It involves skillfully calibrating
between divergent and interpretive capabilities in response to evolving
research needs and contexts.

\subsection{3.3.3 Risk Dynamics}\label{risk-dynamics}

Risk Dynamics represent forces introducing vulnerabilities into
partnerships. If not managed, these dynamics can undermine epistemic
integrity, trust, and collaboration value.

\textbf{Epistemic Alienation.} Epistemic Alienation refers to where
researchers feel dispossessed of interpretive ownership over knowledge
they formally endorse. They may accept AI-generated outputs as valid but
struggle to fully explain, contextualize, or defend them within existing
epistemic frameworks. This dynamic emerges when the gap between
AI-generated outputs and human interpretive capacities exceeds a
critical threshold, often linked to insufficient Interpretive
Intelligence (Section 3.2.1).

Unlike automation complacency (passive reliance) or system opacity
problems (inability to trace causal chains), Epistemic Alienation
denotes deeper normative and existential loss. Researchers feel
disconnected from the \emph{meaning} and \emph{foundations} of
knowledge, even if they understand its provenance.

Epistemic Alienation risk increases with agency configuration. In
Directed Agency, alienation is minimal as humans maintain tight
interpretive control. In Contributory Agency, partial alienation may
emerge around specific AI-initiated contributions. In Partnership
Agency, where knowledge production becomes emergent and less traceable
to individual contributions, alienation risk increases significantly.

When researchers experience Epistemic Alienation, they may retreat from
more advanced agency configurations to regain interpretive control,
potentially sacrificing innovative potential for epistemic security.
Managing this dynamic requires deliberate attention to interpretive
practices, fostering high Interpretive Intelligence, and developing new
explanatory frameworks bridging human and AI epistemologies.

\textbf{Epistemic Closure.} Epistemic Closure refers to narrowing of
perspective through reinforcing feedback loops between humans and AI
systems. This dynamic emerges when partnerships inadvertently reinforce
existing biases (either human or algorithmic) and reduce interpretive
diversity. It is concerning when AI is trained on biased or partial
datasets, or when researchers interact with AI in ways that consistently
confirm their prior commitments.

Preliminary interviews with AI-augmented research teams reveal how
partnerships risk epistemic closure when feedback loops filter out
dissenting perspectives or novel interpretations. Partnerships may
inadvertently create an echo chamber amplifying certain viewpoints while
dismissing alternatives, gradually reducing diversity of theoretical
approaches considered or methodologies employed.

This dynamic poses significant risk to transformative potential of
partnerships, as innovation often requires challenging established
assumptions and considering diverse viewpoints. Managing Epistemic
Closure requires deliberately introducing diverse perspectives (through
team composition, varied data sources, or incorporating AI designed to
play devil's advocate) and establishing mechanisms for periodically
questioning core assumptions and reinforcing loops.

\subsection{3.3.4 Dynamics Interactions and Threshold Effects}\label{dynamics-interactions-and-threshold-effects}

Partnership Dynamics create complex interaction patterns influencing how
collaborations evolve over time, generating both virtuous and vicious
cycles shaping development trajectories. Strong Epistemic Ambidexterity
can mitigate Epistemic Alienation risk by ensuring human understanding
keeps pace with innovation. Conversely, unchecked Transformative
Potential without sufficient Epistemic Ambidexterity or Interpretive
Intelligence may accelerate Epistemic Alienation.

Particularly significant are potential threshold effects occurring when
dynamics reach critical levels or interact synergistically. When
multiple generative dynamics exceed certain thresholds simultaneously,
qualitative transformations in knowledge production---paradigm
shifts---may become possible. When high levels of Recursive Evolution
and Transformative Potential coincide within a partnership exhibiting
strong Epistemic Ambidexterity, rapid transitions to stable Partnership
Agency configurations might occur, enabling breakthrough innovations
reshaping entire fields.

Understanding these dynamic interactions and potential thresholds helps
research teams anticipate and navigate epistemic transformations more
effectively. The framework equips teams with conceptual tools for
identifying potential vulnerabilities (risk of Epistemic Closure) and
leveraging opportunities (harnessing Transformative Potential) as
partnerships evolve.

\section{3.4 Theoretical Implications}\label{theoretical-implications}

The Cognitio Emergens framework transforms how we understand knowledge
creation in the age of advanced AI, with implications for epistemology
and agency. These implications, along with effects on methodology,
organizations, and ethics, extend beyond specific research contexts to
address questions about the nature of human-AI collaboration.

\subsection{3.4.1 Epistemological Implications}\label{epistemological-implications}

Most fundamentally, the framework suggests a shift in epistemology. It
points beyond both individual human cognition and purely social
knowledge creation to a third epistemological space where knowledge
emerges dynamically from human-AI partnerships. This constitutes a form
of distributed cognition that extends beyond current epistemological
categories.

While Thomas Kuhn described how scientific paradigms shift through
crisis and revolution (Kuhn and Hacking 2012), this framework suggests
how paradigms themselves might be systematically explored, evaluated,
and perhaps generated through mature partnerships possessing advanced
dimensional capabilities (especially Divergent, Synthesis, and
Axiological Intelligence).

The framework also highlights questions about the nature of knowledge
itself. If knowledge traditionally requires production, interpretive
ownership, and justification by human knowers, how do we conceptualize
insights emerging from systems where human interpretive capacity may be
outpaced by generative capability (Shanahan 2025)? This challenges us to
rethink what it means to ``know'' in a human-AI context.

\subsection{3.4.2 Agency Implications}\label{agency-implications}

The framework reconceptualizes agency in knowledge creation as
distributed, dynamic, and emergent, rather than singular, fixed, and
located solely in humans (Holter and El-Assady 2024; Watkins et al.
2025). Agency arises from structured interaction between humans and AI,
creating forms of distributed cognition possessing capabilities beyond
those of any individual component. This contrasts with agent-centric
views (Silver and Sutton 2025).

This perspective challenges traditional assumptions about epistemic
authority and responsibility. As agency configurations evolve toward
Partnership, the locus of effective epistemic authority shifts from
individual humans to the human-AI system itself. This raises questions
about accountability, attribution of discovery, and definition of
scientific contribution in an era of advanced AI, touching on governance
issues highlighted in recent studies (Lou et al. 2025; Liu and Shen
2025).

\subsection{3.4.3 Methodological Implications}\label{methodological-implications}

The framework suggests that research methodology itself is not static
but can co-evolve with partnerships. Through dimensional development and
agency transitions, research methods can be transformed, reflecting
multi-loop learning processes (Argyris 1976; Tosey, Visser, and Saunders
2012; Lou et al. 2025). Advanced partnerships don't just apply existing
methodologies more efficiently---they can create new methodological
approaches that transcend current disciplinary boundaries.

This implies a shift in how we evaluate research methods. Instead of
viewing methods as fixed tools, we can see them as evolving capabilities
of partnerships. This allows richer questions: ``In which dimensions
(e.g., Synthesis, Anticipatory) has this partnership's methodological
approach evolved furthest? Where are developmental imbalances? What
novel methodologies might emerge if we strategically cultivate specific
dimensions?''

\subsection{3.4.4 Organizational Implications}\label{organizational-implications}

The framework underscores how organizational contexts shape human-AI
epistemic partnerships, addressing gaps identified in the literature
(Lou et al. 2025; Liu and Shen 2025). Factors such as structural
configurations (e.g., modularity, boundary permeability (Joseph and
Sengul 2024)), coordination mechanisms, incentive systems, and cultural
elements (e.g., risk tolerance, interpretive diversity) influence how
partnerships develop, which dimensions flourish, and what they achieve.

This suggests that optimizing human-AI collaboration requires attention
beyond technological capabilities and individual interactions to
surrounding organizational design and culture. Cultivating principles of
organizational ambidexterity (O'Reilly and Tushman 2013)---balancing
exploration and exploitation, fostering psychological safety for
experimentation, and establishing appropriate governance---becomes
essential for realizing potential of advanced agency configurations and
dimensional capabilities.

\subsection{3.4.5 Ethical Implications}\label{ethical-implications}

The framework surfaces ethical questions. The Axiological Intelligence
dimension suggests that scientific values themselves can evolve through
human-AI interaction, potentially creating new axiological frameworks
that transcend current ethical categories. This raises questions about
how we guide this evolution responsibly.

The tension between innovative potential (especially in Partnership
Agency) and risk of Epistemic Alienation highlights a central ethical
challenge: How do we balance pursuit of transformative discovery with
need to maintain meaningful human understanding, interpretation, and
accountability in knowledge creation (Seeber et al. 2018)? This question
becomes increasingly urgent as AI systems evolve from assistants to
co-authors (Beel, Kan, and Baumgart 2025; Yamada et al. 2025), requiring
us to reconsider epistemic and ethical foundations of scientific
practice.

\section{3.5 Conclusion}\label{conclusion}

The Cognitio Emergens framework offers a theoretical structure for
understanding emergent properties of human-AI epistemic partnerships. By
identifying Agency Configurations, Epistemic Dimensions, and Partnership
Dynamics, the framework provides conceptual tools for analyzing current
partnerships while guiding their future development.

Unlike frameworks focused primarily on technical capabilities or ethical
constraints, Cognitio Emergens explicitly addresses the emergent,
co-creative potential of human-AI collaboration in knowledge production.
It acknowledges both opportunities for epistemic transformation and
challenges that may arise as partnerships evolve.

The framework's integration of visual representation with conceptual
structure creates a tool for both theoretical exploration and practical
application. The capability signatures approach provides a systematic
method for understanding the uneven ways that partnerships develop
across dimensions and agency configurations.

As AI systems evolve from instrumental tools to epistemic partners,
frameworks like Cognitio Emergens become essential for understanding and
guiding this transformation. The framework offers a foundation for
developing partnerships that maximize transformative potential while
maintaining epistemic integrity.

\section{4. Implementing the Cognitio Emergens Framework}\label{implementing-the-cognitio-emergens-framework}

This section bridges theoretical concepts with practical application,
translating the Cognitio Emergens framework into actionable
implementation guidance. We address three implementation domains:
configuring agency relationships (4.1), cultivating epistemic dimensions
(4.2), and managing partnership dynamics (4.3). For each domain, we
provide specific actions and their corresponding value, offering a
flexible foundation that research teams can adapt to their unique
contexts. While Appendix A provides more comprehensive strategies and
assessment protocols, these core principles offer a starting point for
organizations seeking to implement the framework.

\subsection{4.1 Configuring Agency Relationships}\label{configuring-agency-relationships}

Implementing agency configurations---the foundation of the CE
framework---requires deliberate design choices that balance innovation
potential with interpretive control across different research contexts.
Navigating these configurations allows teams to leverage AI
appropriately for different research phases and goals.

For \textbf{Directed Agency}---where AI functions as a tool within
human-defined boundaries---implementation focuses on establishing clear
parameters and robust validation.

\begin{itemize}
\tightlist
\item
  \textbf{Actions:} Define explicit constraints for AI tasks (permitted
  data sources, banned methodologies, required output formats). Develop
  systematic verification protocols aligned with disciplinary standards
  (checklists for result validation, independent replication
  procedures).
\item
  \textbf{Value:} Ensures AI contributions enhance efficiency without
  compromising established epistemic rigor, maintaining clear human
  authority.
\end{itemize}

For \textbf{Contributory Agency}---where AI actively contributes novel
ideas---implementation requires creating receptive spaces while
maintaining evaluative oversight.

\begin{itemize}
\tightlist
\item
  \textbf{Actions:} Schedule dedicated exploratory sessions where
  unprompted AI insights (anomaly detection reports, alternative
  hypothesis suggestions) can be considered based on potential rather
  than immediate fit. Develop evaluation criteria that balance novelty
  with epistemic grounding (rating AI suggestions on plausibility
  \emph{and} originality). Implement workflows where AI suggestions
  trigger specific human review processes.
\item
  \textbf{Value:} Allows teams to benefit from AI's potential to spark
  innovation and challenge assumptions, moving beyond single-loop
  learning without sacrificing critical evaluation.
\end{itemize}

For \textbf{Partnership Agency}---where human-AI boundaries blur into
unified systems---implementation necessitates iterative workflows and
shared interpretive structures.

\begin{itemize}
\tightlist
\item
  \textbf{Actions:} Establish rapid, iterative co-creation workflows
  where human and AI contributions build upon each other with minimal
  delay (using shared digital whiteboards or collaborative coding
  environments). Foster interpretive communities---groups of researchers
  who collectively examine, debate, and interpret partnership
  outputs---to build shared ownership and navigate complexity. Develop
  shared conceptual vocabularies or `partnership dialects' to facilitate
  communication about emergent concepts.
\item
  \textbf{Value:} Maximizes potential for breakthrough insights emerging
  from the synergy of human and AI cognition, enabling triple-loop
  learning and paradigm shifts.
\end{itemize}

As partnerships oscillate between configurations, implementation must
include strategies for managing these transitions effectively.

\begin{itemize}
\tightlist
\item
  \textbf{Actions:} Identify early signals suggesting configuration
  shifts (increasing interpretive challenges signaling a need to
  retreat; breakthrough AI suggestions signaling opportunities to
  advance). Develop team capacity (through training and reflective
  practice) to operate across multiple configurations, consciously
  selecting the mode best suited to the current task (using Partnership
  for brainstorming, Directed for validation).
\item
  \textbf{Value:} Flexible adaptation across configurations allows teams
  to optimize collaboration for different research needs, enhancing
  overall project coherence and effectiveness.
\end{itemize}

While configuring appropriate agency relationships establishes
structural conditions for productive collaboration, cultivating specific
epistemic dimensions develops the partnership's functional capabilities.
These dimensions represent collaborative capacities that emerge through
sustained interaction.

\subsection{4.2 Cultivating Epistemic Dimensions}\label{cultivating-epistemic-dimensions}

Developing the partnership's functional capabilities requires deliberate
cultivation of the six epistemic dimensions---collaborative capacities
that emerge through sustained interaction and targeted development
efforts. This involves fostering specific practices and environments
that encourage development of desired capabilities, leading to more
potent and balanced partnerships.

Along the \textbf{Discovery axis:}

\begin{itemize}
\tightlist
\item
  \textbf{Divergent Intelligence:} Foster by creating psychologically
  safe environments supporting novel hypothesis generation
  (brainstorming sessions where AI outputs are initially treated as
  provocations, not solutions). Implement variation techniques (prompt
  engineering that explicitly asks AI for outlier perspectives or
  counter-arguments) to expand the space of considered possibilities.
  \textbf{Value:} Enhances ability to generate potentially
  groundbreaking ideas.
\item
  \textbf{Interpretive Intelligence:} Develop protocols for generating
  different types of explanations calibrated to specific research needs
  (requesting simplified summaries for quick checks, detailed causal
  trace explanations for critical validation). Utilize interactive
  visualization tools that allow researchers to explore AI models or
  outputs. Implement regular `interpretability audits' where the team
  assesses understandability of recent AI contributions. \textbf{Value:}
  Builds trust, facilitates deeper collaboration, and mitigates the risk
  of epistemic alienation.
\end{itemize}

The \textbf{Integration axis} requires approaches that connect and
synthesize knowledge:

\begin{itemize}
\tightlist
\item
  \textbf{Connective Intelligence:} Implement cross-domain exposure
  practices (tasking AI to explicitly search for analogous problems or
  solutions in adjacent fields; dedicating team time to exploring
  AI-identified cross-disciplinary links). Utilize network visualization
  tools to map connections surfaced by AI across disparate literatures
  or datasets. \textbf{Value:} Facilitates interdisciplinary
  breakthroughs and avoids reinventing wheels.
\item
  \textbf{Synthesis Intelligence:} Organize integration workshops where
  teams explicitly work to combine diverse perspectives (human
  theoretical insights, AI data patterns) into coherent explanatory
  frameworks. Use multi-level modeling approaches capable of
  incorporating findings at different scales or from different
  methodologies. Task AI with generating draft syntheses or conceptual
  maps based on provided inputs. \textbf{Value:} Develops more
  comprehensive understandings by integrating fragmented knowledge.
\end{itemize}

The \textbf{Projection axis} addresses future possibilities and their
significance:

\begin{itemize}
\tightlist
\item
  \textbf{Anticipatory Intelligence:} Develop structured scenario
  exploration approaches (using AI to generate multiple plausible future
  trajectories based on different assumptions; conducting `pre-mortem'
  analyses on potential research pathways). Implement `early warning
  systems' where AI monitors trends for indicators of potential
  disruptions or emerging opportunities. \textbf{Value:} Enables more
  proactive and strategic research planning, especially in fast-moving
  fields.
\item
  \textbf{Axiological Intelligence:} Conduct structured dialogues
  examining the epistemic values embedded in research choices and AI
  outputs (using AI to surface implicit assumptions in assessment
  criteria; team discussions on what constitutes a `significant'
  finding). Co-develop new evaluation criteria with AI that reflect the
  unique contributions of the partnership. \textbf{Value:} Fosters
  critical reflection on research goals and ensures alignment between
  partnership activities and deeper scientific values, potentially
  reshaping disciplinary norms.
\end{itemize}

Implementation approaches should target dimensions strategically based
on the partnership's current capability profile and goals, rather than
attempting uniform advancement. Focusing on dimensions that complement
existing strengths or address weaknesses allows teams to develop
balanced capability signatures tailored to their specific research
contexts, maximizing their collaborative potential.

As epistemic dimensions develop, their evolution is shaped by broader
partnership dynamics that require active management. These dynamics
influence how capabilities evolve over time and determine whether
partnerships achieve their full potential.

\subsection{4.3 Managing Partnership Dynamics}\label{managing-partnership-dynamics}

Managing partnership evolution effectively requires proactive attention
to the generative, balancing, and risk dynamics that shape how
capabilities develop over time and determine whether partnerships
achieve breakthrough potential. This involves cultivating generative
forces, managing balancing tensions, and mitigating risks through
specific practices and continuous monitoring.

\textbf{Generative dynamics} (Transformative Potential, Recursive
Evolution) require deliberate cultivation:

\begin{itemize}
\tightlist
\item
  \textbf{Actions:} Implement practices that periodically challenge
  disciplinary paradigms (`what if' sessions guided by AI provocations).
  Explicitly document the partnership's evolution, including shifts in
  understanding, methodology, or interaction protocols, fostering
  Recursive Evolution. Dedicate resources to exploring high-risk,
  high-reward avenues suggested by the partnership.
\item
  \textbf{Value:} Creates conditions for breakthrough innovation and
  sustained partnership growth beyond incremental improvements.
\end{itemize}

\textbf{Balancing dynamics} (Temporal Integration, Epistemic
Ambidexterity) require conscious management:

\begin{itemize}
\tightlist
\item
  \textbf{Actions:} Establish practices for modulating research pace
  based on epistemic needs (scheduled `slow thinking' sessions for deep
  interpretation; defined `sprints' for rapid exploration). Implement
  clear protocols for switching between exploratory modes (prioritizing
  novelty) and validation modes (prioritizing rigor). Foster Epistemic
  Ambidexterity through regular reflection on the balance between
  innovation and interpretability.
\item
  \textbf{Value:} Enables sustainable progress by managing inherent
  tensions, ensuring both innovation and epistemic integrity are
  maintained.
\end{itemize}

\textbf{Risk dynamics} (Epistemic Alienation, Epistemic Closure) require
preventative and responsive measures:

\begin{itemize}
\tightlist
\item
  \textbf{Actions:} Develop practices for detecting early signs of
  Epistemic Alienation (researcher reluctance to explain AI outputs,
  divergence between formal endorsement and intuitive understanding) and
  intervene (through focused interpretability efforts or temporary
  shifts to more Directed agency). Mitigate Epistemic Closure by
  actively seeking diverse inputs (using multiple AI models, involving
  external reviewers, designing AI prompts to challenge consensus) and
  periodically auditing the partnership for reinforcing biases.
\item
  \textbf{Value:} Protects the partnership's epistemic integrity and
  long-term viability by proactively addressing potential
  vulnerabilities.
\end{itemize}

\subsection{4.4 Conclusion}\label{conclusion-1}

Implementing the Cognitio Emergens framework requires balancing
structured guidance with contextual adaptation. Rather than providing
rigid prescriptions, these implementation principles offer flexible
foundations upon which research teams can build practices appropriate to
their specific contexts, disciplinary traditions, and developmental
goals. Effective implementation is not a one-off setup but an evolving
process requiring ongoing adjustment based on partnership experience and
changing research needs.

When beginning implementation, teams should typically prioritize
configuring appropriate agency relationships before focusing on specific
epistemic dimensions or partnership dynamics. Within agency
configurations, most teams benefit from starting with Directed Agency to
establish shared foundations, gradually incorporating elements of
Contributory Agency as trust and familiarity develop. For epistemic
dimensions, prioritizing Interpretive Intelligence early in the
partnership creates the foundation for other dimensions to flourish.
Each team should assess their unique strengths, vulnerabilities, and
research goals to determine the most appropriate sequence for
implementing various framework components.

Implementation requires appropriate allocation of resources---time for
relationship building and reflective practice, expertise in both domain
knowledge and AI capabilities, and technological infrastructure
supporting collaborative workflows. The degree of investment should
align with the partnership's centrality to research objectives and the
complexity of the collaborative tasks. Initial implementation might
focus on lightweight practices requiring minimal resources, expanding
investment as the partnership demonstrates value.

\section{5. Discussion}\label{discussion}

The Cognitio Emergens framework offers theoretical insight and practical
guidance for navigating human-AI epistemic partnerships. While previous
sections detailed the framework's components and implementation, this
discussion synthesizes its broader significance. We explore the
framework's integrative theoretical contributions, practical
implications for different audiences, directions for future inquiry, and
considerations regarding its scope and boundaries.

\subsection{5.1 Integrative Theoretical Significance}\label{integrative-theoretical-significance}

While section 3.4 detailed specific theoretical implications, here we
examine how these implications collectively shift understanding of
human-AI collaboration. The Cognitio Emergens framework offers five
integrative insights: boundary transcendence, recursive
structure-emergence relationships, constitutive temporal evolution,
theoretical integration, and bridging descriptive-normative
perspectives. Together, these insights reframe how we conceptualize,
study, and guide human-AI epistemic partnerships.

\subsubsection{5.1.1 Beyond Traditional
Boundaries}\label{beyond-traditional-boundaries}

Boundary transcendence moves beyond rigid distinctions between human
knowers and AI tools to reveal a hybrid cognitive system where knowledge
emerges from structural coupling and interaction. This boundary-crossing
manifests simultaneously across epistemology (new forms of distributed
knowing), agency (blurring lines of control and contribution), and
methodology (integrating diverse approaches).

This connects CE to broader intellectual movements challenging dualistic
thinking (ecological approaches to cognition, organizational theories of
permeable boundaries). Just as these movements reconfigured
understanding in their domains, CE reconfigures understanding of
scientific knowledge creation beyond the individual human researcher or
purely social processes.

It situates human-AI collaboration within a larger shift toward
understanding complex phenomena through relationship and interaction.

\subsubsection{5.1.2 Recursive Relationship Between Structure and
Emergence}\label{recursive-relationship-between-structure-and-emergence}

Human-AI partnerships exhibit recursivity: structural configurations
(Agency) and emergent capabilities (Dimensions, Dynamics) mutually
influence each other in ways that challenge linear models of
collaboration. Agency configurations enable capabilities; these
capabilities then reshape agency; dynamics influence both
simultaneously.

This recursive pattern connects CE to complexity theory and studies of
emergence. Human-AI partnerships function as complex adaptive systems.
They generate higher-order properties through component interactions.
These interactions create feedback loops that transform the components
themselves. Guiding these partnerships requires attending to both
structural conditions \emph{and} emergent properties simultaneously,
rather than focusing solely on component design or outcome metrics.

\subsubsection{5.1.3 Temporal Evolution as Constitutive Rather than
Incidental}\label{temporal-evolution-as-constitutive-rather-than-incidental}

CE positions evolution as central to human-AI relationships. Agency
configurations oscillate, dimensions develop unevenly, and dynamics
create evolving cycles. Time and change are inherent.

This temporal orientation connects CE to evolutionary approaches, with a
key difference: CE emphasizes how partnerships \emph{actively construct}
the epistemic environments they inhabit through evolving practices,
rather than just adapting to fixed external conditions. This implies
that sustainable partnerships require deliberate cultivation of
developmental trajectories---balancing adaptation with intentional
shaping of the collaborative process over time.

\subsubsection{5.1.4 Integration of Multiple Theoretical
Traditions}\label{integration-of-multiple-theoretical-traditions}

The framework's power stems partly from its synthesis of multiple
theoretical traditions often treated in isolation (autopoiesis,
organizational design, sociotechnical systems, epistemology, learning
theory). This integration allows CE to address complex phenomena
occurring at the intersection of these domains.

This theoretical integration mirrors the framework's insight about
dimensional capabilities: powerful partnerships integrate capacities
across multiple dimensions. Similarly, understanding these partnerships
benefits from theoretical approaches that integrate insights from
multiple disciplines. CE provides a scaffold for such cross-disciplinary
understanding.

\subsubsection{5.1.5 Bridge Between Descriptive and Normative
Perspectives}\label{bridge-between-descriptive-and-normative-perspectives}

CE bridges descriptive analysis (how partnerships function) and
normative considerations (what constitutes productive collaboration).
While primarily descriptive, its identification of specific dynamics
(balancing, risk) provides a foundation for evaluating partnerships
based on observed patterns rather than purely abstract principles.

This descriptive-normative bridge connects CE to pragmatist traditions
where knowledge informs and transforms practice. It suggests guiding
human-AI partnerships requires continuous oscillation between empirical
observation of \emph{what is} and normative reflection on \emph{what
should be}, allowing each to inform the other.

These integrative insights position Cognitio Emergens as a
reconceptualization of knowledge creation in computationally intensive
environments. By attending to boundary transcendence, recursive
structures, temporal evolution, theoretical integration, and the
descriptive-normative link, the framework provides an approach for
understanding and guiding these emerging epistemic partnerships.

\subsection{5.2 Practical Implications}\label{practical-implications}

The Cognitio Emergens framework addresses the practical tensions
identified in the introduction---particularly the challenges of
theorizing collaboration with blurring boundaries, maintaining
interpretive ownership, and adapting organizational structures. Beyond
resolving these tensions conceptually, the framework delivers concrete
guidance for various stakeholders.

For \textbf{research teams}, the framework provides a shared vocabulary
and diagnostic tools. It enables teams to:

\begin{itemize}
\tightlist
\item \emph{Diagnose their collaborative style:} Identify their current agency configuration(s) and capability signature.
\item \emph{Plan strategically:} Consciously select appropriate agency configurations for different research phases (e.g., Partnership for exploration, Directed for validation).
\item \emph{Identify imbalances:} Recognize potential risks arising from dimensional imbalances (e.g., high Divergence without sufficient Interpretation leading to alienation).
\item \emph{Guide development:} Target specific dimensions or dynamics for cultivation based on project goals.
\end{itemize}

For \textbf{institutional leaders}, CE illuminates crucial
organizational design considerations beyond mere technical
infrastructure. It enables leaders to:

\begin{itemize}
\tightlist
\item \emph{Assess organizational readiness:} Evaluate how current governance structures, evaluation metrics, and culture might enable or constrain different agency configurations.
\item \emph{Foster ambidexterity:} Design organizational structures that support both exploratory Partnership modes (e.g., through dedicated sandboxes or innovation units) and reliable Directed modes.
\item \emph{Adapt incentives:} Reconsider evaluation criteria to recognize and reward contributions emerging from distributed human-AI systems, moving beyond purely individual metrics.
\item \emph{Cultivate culture:} Promote norms that support experimentation, interpretability efforts, and managing risks like epistemic alienation.
\end{itemize}

For \textbf{policymakers and funders}, the framework supplies a more
sophisticated lens for evaluation and support. It enables them to:

\begin{itemize}
\tightlist
\item \emph{Move beyond simple metrics:} Evaluate collaborations based on the development of balanced capability signatures and the management of dynamics, rather than solely on task performance or speed.
\item \emph{Support diverse trajectories:} Recognize the non-linear evolution of partnerships and fund initiatives that support flexible oscillation between agency configurations.
\item \emph{Target funding strategically:} Support the development of specific dimensional capabilities (e.g., Interpretive or Axiological Intelligence) deemed critical for responsible innovation.
\item \emph{Inform ethical guidelines:} Use the framework's concepts (like Epistemic Alienation) to inform policies promoting meaningful human oversight.
\end{itemize}

For \textbf{educators} preparing future researchers, CE suggests
curricula extending beyond technical AI skills. It highlights the need
to:

\begin{itemize}
\tightlist
\item \emph{Develop collaborative literacy:} Train researchers how to effectively engage with AI across different agency configurations.
\item \emph{Cultivate interpretive capacity:} Emphasize skills for understanding, validating, and integrating AI contributions.
\item \emph{Foster critical awareness:} Teach students how to recognize capability signatures, manage partnership dynamics, and navigate the ethical dimensions of human-AI knowledge co-creation.
\end{itemize}

\subsection{5.3 Future Research Directions}\label{future-research-directions}

The Cognitio Emergens framework opens avenues for future research. We
identify five key directions below, with the first two---dynamics of
agency transitions and dimensional interactions---representing urgent
priorities for advancing both theoretical understanding and practical
implementation:

\textbf{Dynamics of Agency Transitions:} How do teams navigate
oscillations between configurations? What triggers shifts or
regressions? Are there patterns characterizing productive transitions
across disciplines? \emph{Expected Contribution:} Deeper understanding
of partnership adaptation mechanisms and best practices for managing
fluidity.

\textbf{Dimensional Interactions and Threshold Effects:} How exactly do
advancements in one dimension enable/constrain others? Do specific
capability signatures predict certain outcomes? Are there identifiable
thresholds where quantitative changes trigger qualitative
transformations in partnership capability? \emph{Expected Contribution:}
Predictive models of partnership evolution and identification of
critical developmental milestones.

\textbf{Sociological and Contextual Factors:} How do power dynamics,
institutional structures, funding mechanisms, and disciplinary cultures
shape partnership evolution, agency configurations, and axiological
negotiations? How do these factors influence management of risks like
Epistemic Closure? \emph{Expected Contribution:} Understanding of how
socio-technical context mediates human-AI collaboration and shapes
outcomes, informing more equitable and context-aware implementations.

\textbf{Sustaining Emergent Capabilities:} What leadership approaches,
organizational designs, routines, and cultural factors enable teams to
sustain advanced capabilities (Partnership agency, high ambidexterity)
over time, resisting regression to simpler modes? \emph{Expected
Contribution:} Evidence-based strategies for embedding and maintaining
high-functioning partnerships within organizations.

\textbf{Operationalizing and Measuring CE Constructs:} How can the
framework's core components (agency configurations, dimensions,
dynamics) be reliably operationalized and measured, both qualitatively
and quantitatively? \emph{Expected Contribution:} Validated assessment
tools for diagnosing partnerships and tracking development
longitudinally.

These research directions would benefit from methodological pluralism,
combining in-depth qualitative studies (ethnography, comparative case
studies) capturing emergent phenomena with quantitative approaches
tracking development across larger samples. Mixed methods designs will
be important for linking observable behaviors with participants'
evolving interpretive frameworks.

\subsection{5.4 Limitations and Boundaries}\label{limitations-and-boundaries}

While these research directions offer pathways for extending the
framework, it is important to recognize its current limitations and
appropriate boundaries of application. These limitations should be
understood not as deficiencies but as defining the framework's scope and
inviting continued theoretical development.

The framework's primary focus on scientific and knowledge-intensive
domains reflects its origins in research contexts where epistemic
outcomes represent the primary objective. Its applicability is more
limited in heavily regulated contexts where interpretive requirements
and accountability considerations necessarily constrain AI autonomy,
such as high-risk medical decision-making or legal adjudication.
Similarly, domains where AI serves primarily operational rather than
epistemic functions will find less relevance in CE's emphasis on
knowledge co-creation.

Nevertheless, the framework's insights about relationship dynamics and
emergent capabilities may offer valuable perspectives even in domains
where full Partnership configurations remain impractical or
inappropriate.

The analytical distinction between epistemic dimensions, while
conceptually useful, inevitably simplifies the complex reality of
emergent epistemic processes. In practice, these dimensions interact and
overlap in ways that resist clean categorization. Divergent and
Interpretive Intelligence often develop in tandem rather than as
separate capabilities; Connective insights enable and are enabled by
Synthesis approaches. This conceptual overlap reflects the messy reality
of knowledge creation but may create challenges for empirical
operationalization. Researchers applying the framework should approach
these dimensions as analytical lenses that illuminate aspects of a
complex phenomenon rather than as cleanly separable categories.

The viability of different agency configurations depends on
organizational and cultural contexts---norms around risk,
interpretability, innovation, and authority shape which configurations
prove productive in particular settings. What constitutes meaningful
Partnership interaction in experimental physics may differ substantially
from its manifestation in clinical medicine or qualitative social
science. The framework should be applied with sensitivity to
disciplinary and organizational contexts rather than assuming universal
manifestations across fields. This context dependence represents not a
limitation in the framework's validity but an acknowledgment of how
epistemic practices remain embedded in specific communities with
distinctive traditions, values, and constraints.

The framework maintains a primarily descriptive rather than prescriptive
orientation, identifying patterns and relationships without universally
privileging any particular agency configuration or dimensional profile.
This reflects its constructivist grounded theory orientation but means
additional normative frameworks may be needed when evaluating ethical
dimensions or determining optimal approaches for specific contexts. The
CE framework itself cannot determine what partnerships should do, only
illuminate how they function and evolve. This normative neutrality
creates space for context-sensitive ethical evaluation but requires
complementary frameworks when making prescriptive judgments about
appropriate configurations for particular domains.

As a new theoretical framework, CE's implementation approaches have not
been extensively validated across diverse research contexts. The
implementation guidance offered represents informed starting points
rather than proven best practices. Organizations implementing the
framework should maintain reflective awareness of how implementation
strategies themselves evolve through experience, treating the framework
as a sensitizing device that guides attention rather than a rigid
template for action.

These limitations define boundaries around CE's appropriate application
while opening spaces for future theoretical refinement through empirical
testing, critical engagement, and interdisciplinary dialogue. They
invite continued development that extends its insights while addressing
its current constraints.

This discussion has illuminated four key aspects of the Cognitio
Emergens framework: its integrative theoretical significance across
multiple traditions, its practical value for diverse stakeholders,
avenues for future research, and its current limitations and boundaries.
The framework's theoretical strength stems from its integration of
boundary transcendence, recursive structure-emergence relationships,
temporal evolution, theoretical synthesis, and bridging of
descriptive-normative perspectives. Its practical applications span
research teams, institutional leaders, policymakers, and educators, each
benefiting from understanding of human-AI epistemic partnerships. The
research directions and acknowledged limitations define a pathway for
the framework's continued development and refinement.

\section{6. Conclusion}\label{conclusion-2}

The Cognitio Emergens framework reconceptualizes human--AI scientific
collaboration as co-evolutionary partnership rather than tool use or
risk management. By integrating agency distribution, emerging
capabilities, and partnership evolution into a coherent model, it
addresses limitations in existing approaches.

Its central insight---that productive partnerships emerge through
co-evolution---extends beyond research to other knowledge-intensive
domains. Future empirical work will refine this framework, but CE
provides a starting point for understanding how humans and AI create
knowledge together. As AI systems become partners in knowledge creation,
we need frameworks capturing the relational, emergent nature of
collaboration. The framework thus invites us: to understand these
partnerships more fully, to guide them more wisely, and to cultivate
partnerships that advance knowledge in ways neither humans nor AI could
achieve alone.

\clearpage

\section{References}\label{references}
\protect\phantomsection\label{refs}

\clearpage

\section{Appendix A: Detailed Implementation Guidance for Cognitio Emergens}\label{appendix-a-detailed-implementation-guidance-for-cognitio-emergens}

This appendix provides comprehensive implementation strategies,
assessment protocols, and reflective questions to supplement the core
principles presented in Section 4. The guidance aims to bridge the gap
between conceptual understanding and practical application, ensuring the
framework's potential is realized effectively. It is organized to
parallel the framework's structure, offering detailed approaches for
each component.

\begin{quote}
\textbf{HOW TO USE THIS APPENDIX}

This appendix is designed as a practical resource that can be
selectively applied based on your specific research context. Each
section includes:

\begin{itemize}
\tightlist
\item
  \textbf{Summary overviews} - Quick reference guides to key concepts
\item
  \textbf{Strategy tables} - Detailed implementation approaches with
  expected benefits
\item
  \textbf{Reflective questions} - Prompts to guide adaptation to your
  context
\end{itemize}

Rather than attempting to implement everything at once, consider which
components best address your current research challenges and
organizational readiness.
\end{quote}

\subsection{A.1 Agency Configuration Implementation}\label{a.1-agency-configuration-implementation}

\begin{quote}
\textbf{AT A GLANCE: AGENCY CONFIGURATIONS}

The framework offers three distinct configurations for human-AI
collaboration, each with different balances of agency:

\begin{itemize}
\tightlist
\item
  \textbf{Directed Agency}: AI as tool within human-defined boundaries
\item
  \textbf{Contributory Agency}: AI actively contributes novel ideas with
  human oversight
\item
  \textbf{Partnership Agency}: Human-AI boundaries blur into unified
  systems
\end{itemize}

Effective implementation involves designing appropriate workflows,
verification methods, and interpretive practices for each configuration.
\end{quote}

\subsubsection{A.1.1 Directed Agency: Detailed
Strategies}\label{a.1.1-directed-agency-detailed-strategies}

\begin{quote}
\textbf{AT A GLANCE: DIRECTED AGENCY}\\
\emph{Deploying AI as a tool within clearly defined human-established
boundaries}

\textbf{Key Components:}

\begin{itemize}
\tightlist
\item
  \textbf{Boundary Specification}: Defining clear parameters for AI
  operation
\item
  \textbf{Verification Protocols}: Ensuring reliability of AI outputs
\item
  \textbf{Interpretive Frameworks}: Translating outputs into
  disciplinary knowledge
\item
  \textbf{Documentation Practices}: Recording processes, decisions, and
  rationales
\end{itemize}
\end{quote}

\paragraph{A.1.1.1 Boundary Specification
Strategies}\label{a.1.1.1-boundary-specification-strategies}

To ensure boundaries are clear, consistent, documented, and adaptable as
partnerships evolve, consider the following strategies:

\begin{longtable}[]{@{}
  >{\raggedright\arraybackslash}p{(\columnwidth - 4\tabcolsep) * \real{0.1786}}
  >{\raggedright\arraybackslash}p{(\columnwidth - 4\tabcolsep) * \real{0.4821}}
  >{\raggedright\arraybackslash}p{(\columnwidth - 4\tabcolsep) * \real{0.3393}}@{}}
\toprule\noalign{}
\begin{minipage}[b]{\linewidth}\raggedright
Strategy
\end{minipage} & \begin{minipage}[b]{\linewidth}\raggedright
Implementation Approaches
\end{minipage} & \begin{minipage}[b]{\linewidth}\raggedright
Expected Benefits
\end{minipage} \\
\midrule\noalign{}
\endhead
\bottomrule\noalign{}
\endlastfoot
\textbf{Documentation Templates} & Specify acceptable data sources and
preprocessing requirements; Define permitted algorithmic approaches and
parameter constraints; Establish required output formats and metadata;
Document verification protocols and acceptance criteria & Creates
consistency across projects; reduces ambiguity; supports onboarding of
new team members; enables systematic boundary evolution \\
\textbf{Parameter Libraries} & Develop repositories of validated
parameter settings; Enable selection from pre-approved configurations;
Include context information for appropriate use cases; Document
performance characteristics & Reduces redundant work; builds on proven
approaches; enables faster configuration; supports knowledge transfer \\
\textbf{Boundary Evolution Protocols} & Establish procedures for
systematically expanding boundaries; Document boundary changes and
rationales; Create graduated permission levels; Implement review
processes for boundary adjustments & Supports partnership maturation;
ensures boundaries evolve deliberately; maintains appropriate
constraints while enabling growth \\
\end{longtable}

\paragraph{A.1.1.2 Verification Protocol
Strategies}\label{a.1.1.2-verification-protocol-strategies}

To ensure the reliability and validity of AI contributions within
defined boundaries, implement robust verification protocols using
strategies such as:

\begin{longtable}[]{@{}
  >{\raggedright\arraybackslash}p{(\columnwidth - 4\tabcolsep) * \real{0.1786}}
  >{\raggedright\arraybackslash}p{(\columnwidth - 4\tabcolsep) * \real{0.4821}}
  >{\raggedright\arraybackslash}p{(\columnwidth - 4\tabcolsep) * \real{0.3393}}@{}}
\toprule\noalign{}
\begin{minipage}[b]{\linewidth}\raggedright
Strategy
\end{minipage} & \begin{minipage}[b]{\linewidth}\raggedright
Implementation Approaches
\end{minipage} & \begin{minipage}[b]{\linewidth}\raggedright
Expected Benefits
\end{minipage} \\
\midrule\noalign{}
\endhead
\bottomrule\noalign{}
\endlastfoot
\textbf{Multi-Method Verification} & Cross-reference with established
literature; Verify against experimental data; Triangulate with
alternative computational methods; Utilize human expert evaluation &
Increases confidence in results; detects potential errors; reduces
reliance on single verification method; balances computational and human
judgment \\
\textbf{Reproducibility Frameworks} & Document and archive input data;
Version and document parameters; Track environmental variables; Analyze
output variance & Supports scientific rigor; enables independent
verification; ensures consistent results; facilitates troubleshooting \\
\textbf{Graduated Verification} & Implement tiered protocols based on
output significance; Apply rigorous verification to outputs that
contradict understanding; Intensify verification for critical research
decisions; Scale verification to output impact & Allocates verification
resources efficiently; focuses scrutiny where needed most; balances
thoroughness with practicality \\
\end{longtable}

\paragraph{A.1.1.3 Interpretive Framework
Strategies}\label{a.1.1.3-interpretive-framework-strategies}

To facilitate the integration of AI outputs into disciplinary knowledge
and ensure human understanding guides the process, consider these
strategies:

\begin{longtable}[]{@{}
  >{\raggedright\arraybackslash}p{(\columnwidth - 4\tabcolsep) * \real{0.1786}}
  >{\raggedright\arraybackslash}p{(\columnwidth - 4\tabcolsep) * \real{0.4821}}
  >{\raggedright\arraybackslash}p{(\columnwidth - 4\tabcolsep) * \real{0.3393}}@{}}
\toprule\noalign{}
\begin{minipage}[b]{\linewidth}\raggedright
Strategy
\end{minipage} & \begin{minipage}[b]{\linewidth}\raggedright
Implementation Approaches
\end{minipage} & \begin{minipage}[b]{\linewidth}\raggedright
Expected Benefits
\end{minipage} \\
\midrule\noalign{}
\endhead
\bottomrule\noalign{}
\endlastfoot
\textbf{Discipline-Specific Interpretation Guides} & Develop
interpretation frameworks for specific fields; Translate AI outputs into
discipline-appropriate concepts; Map computational approaches to
disciplinary methods; Align AI capabilities with domain knowledge needs
& Bridges computational and disciplinary understanding; increases
relevance of AI contributions; builds ownership among domain experts \\
\textbf{Explanation Templates} & Create standardized formats for
expressing AI contributions; Include appropriate qualifications and
context; Develop templates for different output types; Structure
templates to align with disciplinary conventions & Supports consistent
communication; improves clarity; facilitates integration with existing
knowledge; reduces misinterpretation \\
\textbf{Interpretive Workshops} & Establish regular collaborative
interpretation sessions; Include diverse team perspectives; Build shared
understanding of translation approaches; Document interpretation
decisions and rationales & Builds interpretive capacity; surfaces
different perspectives; creates shared ownership; develops team
interpretive capabilities \\
\end{longtable}

\paragraph{A.1.1.4 Documentation Practice
Strategies}\label{a.1.1.4-documentation-practice-strategies}

To maintain transparency, enable reproducibility, and support future
analysis regarding how boundaries were set and respected, implement
rigorous documentation practices:

\begin{longtable}[]{@{}
  >{\raggedright\arraybackslash}p{(\columnwidth - 4\tabcolsep) * \real{0.1786}}
  >{\raggedright\arraybackslash}p{(\columnwidth - 4\tabcolsep) * \real{0.4821}}
  >{\raggedright\arraybackslash}p{(\columnwidth - 4\tabcolsep) * \real{0.3393}}@{}}
\toprule\noalign{}
\begin{minipage}[b]{\linewidth}\raggedright
Strategy
\end{minipage} & \begin{minipage}[b]{\linewidth}\raggedright
Implementation Approaches
\end{minipage} & \begin{minipage}[b]{\linewidth}\raggedright
Expected Benefits
\end{minipage} \\
\midrule\noalign{}
\endhead
\bottomrule\noalign{}
\endlastfoot
\textbf{Input-Process-Output Documentation} & Record input data sources
and preprocessing steps; Document AI configuration details and
parameters; Maintain complete output records including intermediate
states; Track human evaluation and modification processes & Creates
comprehensive audit trail; enables reproduction; supports
troubleshooting; facilitates knowledge transfer \\
\textbf{Provenance Tracking} & Establish origin of all data elements;
Document specific AI systems and versions employed; Record human
involvement in each research stage; Track modification history
throughout the process & Clarifies contributions; supports attribution;
enables tracing of influence; supports accountability \\
\textbf{Accessibility-Oriented Documentation} & Create documentation for
multiple stakeholder perspectives; Develop versions for different
technical expertise levels; Design for potential temporal distance from
research; Structure to support reanalysis needs & Increases usability;
supports diverse audiences; enables long-term utility; facilitates
external review \\
\end{longtable}

\begin{center}\rule{0.5\linewidth}{0.5pt}\end{center}

\begin{quote}
\textbf{REFLECTIVE QUESTIONS: DIRECTED AGENCY}

Consider these questions to strengthen your approach to Directed Agency:

\begin{enumerate}
\def\labelenumi{\arabic{enumi}.}
\item
  Reflecting on your project's goals and risks, which verification
  standards (from A.1.1.2) offer the best balance of rigor and
  feasibility, and why?
\item
  How might standardizing boundaries (A.1.1.1) support consistency, and
  what steps can mitigate the risk of unduly constraining useful
  diversity in AI contributions?
\item
  Which interpretive frameworks (A.1.1.3) best align with your
  disciplinary traditions while effectively accommodating computational
  approaches?
\item
  What specific mechanisms can ensure AI contributions remain within
  established boundaries (A.1.1.1) without unnecessarily filtering
  potentially useful outputs?
\item
  Considering potential future uses (reanalysis, audits), which
  documentation approaches (A.1.1.4) most effectively capture both the
  boundaries themselves and their underlying rationales?
\end{enumerate}
\end{quote}

\subsubsection{A.1.2 Contributory Agency: Detailed
Strategies}\label{a.1.2-contributory-agency-detailed-strategies}

\begin{quote}
\textbf{AT A GLANCE: CONTRIBUTORY AGENCY}\\
\emph{Creating conditions where AI can introduce novel ideas while
humans maintain evaluative oversight}

\textbf{Key Components:}

\begin{itemize}
\tightlist
\item
  \textbf{Receptive Spaces:} Environments for considering AI-initiated
  suggestions
\item
  \textbf{Evaluation Frameworks:} Systems for assessing potential value
  of AI contributions
\item
  \textbf{Influence Documentation:} Methods for tracking idea provenance
  and attribution
\item
  \textbf{Reflexive Practice:} Approaches for maintaining critical
  awareness
\end{itemize}
\end{quote}

\paragraph{A.1.2.1 Receptive Spaces
Strategies}\label{a.1.2.1-receptive-spaces-strategies}

To create environments where novel AI contributions can be surfaced and
considered seriously, even if initially unconventional, explore these
strategies:

\begin{longtable}[]{@{}
  >{\raggedright\arraybackslash}p{(\columnwidth - 4\tabcolsep) * \real{0.1786}}
  >{\raggedright\arraybackslash}p{(\columnwidth - 4\tabcolsep) * \real{0.4821}}
  >{\raggedright\arraybackslash}p{(\columnwidth - 4\tabcolsep) * \real{0.3393}}@{}}
\toprule\noalign{}
\begin{minipage}[b]{\linewidth}\raggedright
Strategy
\end{minipage} & \begin{minipage}[b]{\linewidth}\raggedright
Implementation Approaches
\end{minipage} & \begin{minipage}[b]{\linewidth}\raggedright
Expected Benefits
\end{minipage} \\
\midrule\noalign{}
\endhead
\bottomrule\noalign{}
\endlastfoot
\textbf{Dedicated Exploration Sessions} & Begin with divergent
consideration of multiple AI contributions; Progress to evaluation of
promising suggestions; Conclude with integration of selected
contributions into research plans & Creates protected time for novel
ideas; reduces pressure for immediate application; establishes routine
for considering AI-initiated directions \\
\textbf{Contribution Repositories} & Store original AI contributions
with timestamps and contexts; Include initial human assessments and
potential applications; Tag for retrieval based on research needs;
Implement periodic review mechanisms & Preserves potentially valuable
ideas for future consideration; prevents premature dismissal; enables
pattern recognition across suggestions over time \\
\textbf{Contextual Priming Approaches} & Expose AI to diverse literature
and data sources; Intentionally relax problem constraints;
Systematically vary prompting approaches; Temporarily suspend
disciplinary assumptions & Creates fertile conditions for novel AI
suggestions; increases diversity of contributions; helps AI transcend
training limitations \\
\end{longtable}

\paragraph{A.1.2.2 Evaluation Framework
Strategies}\label{a.1.2.2-evaluation-framework-strategies}

To systematically assess the potential value of AI-initiated
contributions beyond immediate applicability, develop structured
evaluation frameworks:

\begin{longtable}[]{@{}
  >{\raggedright\arraybackslash}p{(\columnwidth - 4\tabcolsep) * \real{0.1786}}
  >{\raggedright\arraybackslash}p{(\columnwidth - 4\tabcolsep) * \real{0.4821}}
  >{\raggedright\arraybackslash}p{(\columnwidth - 4\tabcolsep) * \real{0.3393}}@{}}
\toprule\noalign{}
\begin{minipage}[b]{\linewidth}\raggedright
Strategy
\end{minipage} & \begin{minipage}[b]{\linewidth}\raggedright
Implementation Approaches
\end{minipage} & \begin{minipage}[b]{\linewidth}\raggedright
Expected Benefits
\end{minipage} \\
\midrule\noalign{}
\endhead
\bottomrule\noalign{}
\endlastfoot
\textbf{Multi-Criteria Evaluation Matrices} & Assess novelty
(distinctiveness from existing approaches); Evaluate plausibility
(alignment with established understanding); Consider fertility
(potential to generate new directions); Analyze feasibility (practical
implementability); Judge significance (potential impact if successful) &
Balances multiple dimensions of value; prevents dismissal based on
single criterion; provides structure for fair assessment \\
\textbf{Staged Evaluation Protocols} & Initial screening based on basic
plausibility; Secondary assessment of novelty and potential
significance; Detailed evaluation of methodological feasibility; Final
assessment of integration potential & Efficiently allocates attention;
applies increasing scrutiny to promising ideas; avoids premature
filtering \\
\textbf{Comparative Evaluation Approaches} & Compare to human-generated
alternatives; Contrast with previous AI contributions; Benchmark against
established approaches; Compare contributions from different AI systems
& Provides contextual reference points; highlights distinctive value;
reduces novelty bias \\
\end{longtable}

\paragraph{A.1.2.3 Influence Documentation
Strategies}\label{a.1.2.3-influence-documentation-strategies}

To track the provenance of ideas and appropriately acknowledge AI's
contributory role in research outputs, implement clear documentation
methods:

\begin{longtable}[]{@{}
  >{\raggedright\arraybackslash}p{(\columnwidth - 4\tabcolsep) * \real{0.1786}}
  >{\raggedright\arraybackslash}p{(\columnwidth - 4\tabcolsep) * \real{0.4821}}
  >{\raggedright\arraybackslash}p{(\columnwidth - 4\tabcolsep) * \real{0.3393}}@{}}
\toprule\noalign{}
\begin{minipage}[b]{\linewidth}\raggedright
Strategy
\end{minipage} & \begin{minipage}[b]{\linewidth}\raggedright
Implementation Approaches
\end{minipage} & \begin{minipage}[b]{\linewidth}\raggedright
Expected Benefits
\end{minipage} \\
\midrule\noalign{}
\endhead
\bottomrule\noalign{}
\endlastfoot
\textbf{Contribution Genealogies} & Document initial AI suggestions and
contexts; Track human modifications and extensions; Record subsequent AI
refinements; Capture final implemented versions & Visualizes
evolutionary path of ideas; clarifies human-AI interplay; supports
appropriate attribution \\
\textbf{Attribution Protocols} & Develop internal documentation for team
reference; Create external attribution in publications; Implement
metadata tagging for digital research objects; Establish citation
practices for significant AI contributions & Ensures transparency;
supports ethical acknowledgment; enables tracing of influence \\
\textbf{Influence Assessment} & Map research path changes following AI
suggestions; Identify conceptual shifts initiated by AI; Document
methodological innovations triggered by AI; Assess how AI contributions
altered research priorities & Reveals deeper patterns of influence;
supports meta-reflection; informs future collaboration strategies \\
\end{longtable}

\paragraph{A.1.2.4 Reflexive Practice
Strategies}\label{a.1.2.4-reflexive-practice-strategies}

To cultivate awareness of biases and maintain appropriate critical
distance when evaluating AI contributions, foster reflexive practices
among researchers:

\begin{longtable}[]{@{}
  >{\raggedright\arraybackslash}p{(\columnwidth - 4\tabcolsep) * \real{0.1786}}
  >{\raggedright\arraybackslash}p{(\columnwidth - 4\tabcolsep) * \real{0.4821}}
  >{\raggedright\arraybackslash}p{(\columnwidth - 4\tabcolsep) * \real{0.3393}}@{}}
\toprule\noalign{}
\begin{minipage}[b]{\linewidth}\raggedright
Strategy
\end{minipage} & \begin{minipage}[b]{\linewidth}\raggedright
Implementation Approaches
\end{minipage} & \begin{minipage}[b]{\linewidth}\raggedright
Expected Benefits
\end{minipage} \\
\midrule\noalign{}
\endhead
\bottomrule\noalign{}
\endlastfoot
\textbf{Contribution Journals} & Record initial reactions to AI
contributions; Document evolution of perspective on AI suggestions;
Reflect on contributions initially deemed unproductive; Identify
patterns in receptivity & Builds self-awareness; reveals biases;
encourages reconsideration of dismissed ideas \\
\textbf{Analytical Distance Techniques} & Implement structured
skepticism protocols; Develop contribution-independent evaluation
criteria; Conduct comparative analysis with non-AI approaches;
Explicitly identify implicit acceptance biases & Maintains critical
stance; prevents uncritical acceptance; supports balanced assessment \\
\textbf{Epistemic Evolution Tracking} & Document terminology evolution;
Track conceptual framework modifications; Record methodological approach
transformations; Monitor evaluation criteria developments & Captures
subtle shifts in thinking; reveals AI influence on frameworks; supports
meta-cognitive awareness \\
\end{longtable}

\begin{center}\rule{0.5\linewidth}{0.5pt}\end{center}

\begin{quote}
\textbf{REFLECTIVE QUESTIONS: CONTRIBUTORY AGENCY}

Consider these questions to strengthen your team's approach to
Contributory Agency:

\begin{enumerate}
\def\labelenumi{\arabic{enumi}.}
\item
  How can your team balance receptivity to novel AI contributions
  (cultivated via A.1.2.1) with maintaining research focus and
  coherence?
\item
  Considering your field's epistemic standards, what evaluation criteria
  (A.1.2.2) appropriately assess AI-initiated insights without imposing
  excessive disciplinary constraints or dismissing novelty?
\item
  How might documentation practices (A.1.2.3) best capture the evolving
  nature of human-AI influence without oversimplifying complex
  interactions or creating undue burden?
\item
  What attribution models (inspired by A.1.2.3) most ethically and
  accurately represent the contributory nature of AI in your research
  outputs while maintaining appropriate human responsibility?
\item
  How can exploratory spaces (A.1.2.1) be structured to encourage
  useful, novel AI contributions without sacrificing overall research
  productivity?
\end{enumerate}
\end{quote}

\subsubsection{A.1.3 Partnership Agency: Detailed
Strategies}\label{a.1.3-partnership-agency-detailed-strategies}

\begin{quote}
\textbf{AT A GLANCE: PARTNERSHIP AGENCY}\\
\emph{Creating integrated human-AI systems where boundaries blur and
emergent insights arise from deep collaboration}

\textbf{Key Components:}

\begin{itemize}
\tightlist
\item
  \textbf{Co-Creation Processes}: Enabling fluid, iterative interaction
\item
  \textbf{Interpretive Communities}: Supporting collective sense-making
\item
  \textbf{Reflexive Boundaries}: Monitoring and adjusting constraints
  dynamically
\item
  \textbf{Shared Conceptual Spaces}: Developing mutual language and
  understanding
\end{itemize}
\end{quote}

\paragraph{A.1.3.1 Co-Creation Process
Strategies}\label{a.1.3.1-co-creation-process-strategies}

To enable the fluid, synergistic interaction characteristic of
Partnership Agency, design processes and environments that support rapid
iteration and emergent idea capture:

\begin{longtable}[]{@{}
  >{\raggedright\arraybackslash}p{(\columnwidth - 4\tabcolsep) * \real{0.1786}}
  >{\raggedright\arraybackslash}p{(\columnwidth - 4\tabcolsep) * \real{0.4821}}
  >{\raggedright\arraybackslash}p{(\columnwidth - 4\tabcolsep) * \real{0.3393}}@{}}
\toprule\noalign{}
\begin{minipage}[b]{\linewidth}\raggedright
Strategy
\end{minipage} & \begin{minipage}[b]{\linewidth}\raggedright
Implementation Approaches
\end{minipage} & \begin{minipage}[b]{\linewidth}\raggedright
Expected Benefits
\end{minipage} \\
\midrule\noalign{}
\endhead
\bottomrule\noalign{}
\endlastfoot
\textbf{Continuous Interaction Environments} & Implement real-time
feedback mechanisms; Deploy continuous learning systems that evolve
through interaction; Maintain persistent contextual awareness across
sessions; Develop multi-modal interaction capabilities & Reduces
friction between ideation and implementation; enables rapid evolution of
ideas; supports seamless exchange; fosters deep collaboration \\
\textbf{Iteration Acceleration} & Create streamlined feedback
mechanisms; Build rapid prototyping capabilities; Enable parallel
exploration of multiple trajectories; Establish progressive refinement
protocols & Shortens idea development cycles; increases exploration
breadth; supports rapid testing and refinement; builds momentum \\
\textbf{Emergent Capture} & Implement session recording with interaction
timestamps; Track conceptual evolution through exchanges; Develop
distributed attribution approaches; Identify patterns in emergence &
Documents ideas that neither partner could claim individually; preserves
the developmental history; supports attribution; reveals emergence
patterns \\
\end{longtable}

\paragraph{A.1.3.2 Interpretive Community
Strategies}\label{a.1.3.2-interpretive-community-strategies}

To manage the complexity and potential ambiguity of partnership outputs,
establish shared interpretive practices that foster collective
understanding and ownership:

\begin{longtable}[]{@{}
  >{\raggedright\arraybackslash}p{(\columnwidth - 4\tabcolsep) * \real{0.1786}}
  >{\raggedright\arraybackslash}p{(\columnwidth - 4\tabcolsep) * \real{0.4821}}
  >{\raggedright\arraybackslash}p{(\columnwidth - 4\tabcolsep) * \real{0.3393}}@{}}
\toprule\noalign{}
\begin{minipage}[b]{\linewidth}\raggedright
Strategy
\end{minipage} & \begin{minipage}[b]{\linewidth}\raggedright
Implementation Approaches
\end{minipage} & \begin{minipage}[b]{\linewidth}\raggedright
Expected Benefits
\end{minipage} \\
\midrule\noalign{}
\endhead
\bottomrule\noalign{}
\endlastfoot
\textbf{Collaborative Interpretation Sessions} & Include diverse
disciplinary perspectives; Apply multiple interpretive frameworks;
Implement structured disagreement protocols; Develop consensus-building
mechanisms & Enriches interpretation with multiple viewpoints; surfaces
hidden assumptions; converts disagreement into insight; builds shared
ownership \\
\textbf{Interpretation Rotation} & Designate primary interpreters for
specific outputs; Assign secondary reviewers for alternative
perspectives; Create cross-disciplinary interpretation pairings;
Separate interpretation rounds temporally & Prevents interpretive
monopolies; enriches understanding through diverse perspectives; reduces
individual bias; enables fresh insights \\
\textbf{Interpretive Pluralism} & Create multi-framework analysis
templates; Document competing interpretations systematically; Develop
synthesis approaches for diverse interpretations; Establish protocols
for managing interpretive tensions & Maintains richness of
understanding; prevents premature convergence; supports nuanced
interpretation; leverages productive tension \\
\end{longtable}

\paragraph{A.1.3.3 Reflexive Boundary
Strategies}\label{a.1.3.3-reflexive-boundary-strategies}

To manage the increased autonomy in Partnership Agency responsibly,
implement mechanisms for monitoring and adjusting boundaries dynamically
based on emerging risks or concerns:

\begin{longtable}[]{@{}
  >{\raggedright\arraybackslash}p{(\columnwidth - 4\tabcolsep) * \real{0.1786}}
  >{\raggedright\arraybackslash}p{(\columnwidth - 4\tabcolsep) * \real{0.4821}}
  >{\raggedright\arraybackslash}p{(\columnwidth - 4\tabcolsep) * \real{0.3393}}@{}}
\toprule\noalign{}
\begin{minipage}[b]{\linewidth}\raggedright
Strategy
\end{minipage} & \begin{minipage}[b]{\linewidth}\raggedright
Implementation Approaches
\end{minipage} & \begin{minipage}[b]{\linewidth}\raggedright
Expected Benefits
\end{minipage} \\
\midrule\noalign{}
\endhead
\bottomrule\noalign{}
\endlastfoot
\textbf{Boundary Notification Systems} & Monitor for ethical boundary
approaches; Implement epistemological risk detection; Conduct
methodological validity checks; Assess practical feasibility
continuously & Provides early warning of potential issues; maintains
appropriate guardrails; balances freedom with responsibility; prevents
harmful trajectories \\
\textbf{Temporary Constraint Protocols} & Define targeted constraints
for specific concerns; Implement constraint mechanisms with clear
purpose; Specify constraint durations; Establish release criteria &
Addresses specific risks without permanent restrictions; maintains
partnership agency where appropriate; provides clear rationales for
constraints \\
\textbf{Boundary Reflection} & Document boundary implementations
systematically; Assess constraint effects on partnership outcomes;
Identify alternative approaches to similar concerns; Integrate learning
for future boundary decisions & Builds understanding of boundary
impacts; improves constraint design; supports evolution of boundary
practices; creates organizational learning \\
\end{longtable}

\paragraph{A.1.3.4 Shared Conceptual Space
Strategies}\label{a.1.3.4-shared-conceptual-space-strategies}

To facilitate deep co-creation and mutual understanding, cultivate
shared language and representations that evolve with the partnership:

\begin{longtable}[]{@{}
  >{\raggedright\arraybackslash}p{(\columnwidth - 4\tabcolsep) * \real{0.1786}}
  >{\raggedright\arraybackslash}p{(\columnwidth - 4\tabcolsep) * \real{0.4821}}
  >{\raggedright\arraybackslash}p{(\columnwidth - 4\tabcolsep) * \real{0.3393}}@{}}
\toprule\noalign{}
\begin{minipage}[b]{\linewidth}\raggedright
Strategy
\end{minipage} & \begin{minipage}[b]{\linewidth}\raggedright
Implementation Approaches
\end{minipage} & \begin{minipage}[b]{\linewidth}\raggedright
Expected Benefits
\end{minipage} \\
\midrule\noalign{}
\endhead
\bottomrule\noalign{}
\endlastfoot
\textbf{Vocabulary Evolution} & Document emergent terms and concepts;
Establish meaning negotiation protocols; Track terminological evolution;
Assess conceptual stability & Supports communication about novel
concepts; creates shared language; documents conceptual development;
enables stable reference \\
\textbf{Representational Co-Development} & Implement collaborative
visualization approaches; Create joint modeling techniques; Establish
shared diagramming practices; Track co-evolutionary representation
changes & Builds shared mental models; enables complex idea
communication; supports mutual understanding; creates artifacts
reflecting partnership thinking \\
\textbf{Epistemic Scaffolding} & Develop progressive framework building
techniques; Create conceptual bridge development processes; Implement
transitional vocabulary approaches; Cultivate integrative knowledge
structures & Supports development of novel concepts; connects new ideas
to existing understanding; enables evolution of thought; facilitates
knowledge integration \\
\end{longtable}

\begin{center}\rule{0.5\linewidth}{0.5pt}\end{center}

\begin{quote}
\textbf{REFLECTIVE QUESTIONS: PARTNERSHIP AGENCY}

Consider these questions to strengthen your approach to Partnership
Agency:

\begin{enumerate}
\def\labelenumi{\arabic{enumi}.}
\item
  What iterative co-creation processes (A.1.3.1) best support synergy in
  your research context without creating excessive coordination overhead
  or losing track of emergent ideas?
\item
  How might shared conceptual frameworks (A.1.3.4) emerge organically
  through partnership without being prematurely imposed by either human
  disciplinary norms or AI patterns?
\item
  What approaches to interpretation (A.1.3.2) maintain meaningful human
  participation and critical oversight without unnecessarily
  constraining potentially valuable emergent insights?
\item
  How can reflexive boundaries (A.1.3.3) address ethical and
  epistemological concerns dynamically without permanently restricting
  the partnership's generative potential?
\item
  What specific indicators (qualitative or quantitative) might signal
  when partnership boundaries (A.1.3.3) need temporary strengthening or
  relaxation?
\end{enumerate}
\end{quote}

\subsubsection{A.1.4 Transition Management: Detailed
Strategies}\label{a.1.4-transition-management-detailed-strategies}

\begin{quote}
\textbf{AT A GLANCE: TRANSITION MANAGEMENT}\\
\emph{Developing capabilities to move flexibly between different agency
configurations as research needs evolve}

\textbf{Key Components:}

-\textbf{Transition Indicators}: Recognizing signals suggesting
beneficial configuration changes - \textbf{Configuration Flexibility}:
Building capacity to operate in multiple modes - \textbf{Transition
Protocols}: Managing configuration shifts effectively
\end{quote}

\paragraph{A.1.4.1 Transition Indicator
Strategies}\label{a.1.4.1-transition-indicator-strategies}

Recognizing when a shift between agency configurations might be
beneficial requires systematic monitoring. Develop indicator frameworks
to identify signals suggesting readiness for transitions:

\begin{longtable}[]{@{}
  >{\raggedright\arraybackslash}p{(\columnwidth - 4\tabcolsep) * \real{0.1786}}
  >{\raggedright\arraybackslash}p{(\columnwidth - 4\tabcolsep) * \real{0.4821}}
  >{\raggedright\arraybackslash}p{(\columnwidth - 4\tabcolsep) * \real{0.3393}}@{}}
\toprule\noalign{}
\begin{minipage}[b]{\linewidth}\raggedright
Strategy
\end{minipage} & \begin{minipage}[b]{\linewidth}\raggedright
Implementation Approaches
\end{minipage} & \begin{minipage}[b]{\linewidth}\raggedright
Expected Benefits
\end{minipage} \\
\midrule\noalign{}
\endhead
\bottomrule\noalign{}
\endlastfoot
\textbf{Indicator Frameworks} & \textbf{Directed to Contributory}:
Monitor for increasing comfort with AI outputs, growing evaluation
capacity, emerging AI-identified insights, robust verification
protocols; \textbf{Contributory to Partnership}: Watch for blurring of
contribution boundaries, shared vocabularies, emergent insights neither
could articulate alone, established mutual trust; \textbf{Partnership to
Contributory}: Monitor for epistemic alienation, need for increased
interpretive control, attribution requirements, policy constraints;
\textbf{Contributory to Directed}: Recognize critical phases requiring
maximum reliability, verification challenges, standardization needs,
external scrutiny & Enables timely recognition of transition
opportunities; supports proactive configuration shifts; aligns
configurations with evolving needs; prevents configuration mismatch \\
\textbf{Transition Assessment Tools} & Develop team capability
assessment instruments; Implement epistemic risk evaluation approaches;
Assess organizational readiness; Evaluate project phase appropriateness
& Provides structured evaluation of transition readiness; identifies
capability gaps; surfaces organizational constraints; aligns with
project needs \\
\textbf{Early Warning Systems} & Conduct regular configuration fitness
evaluations; Monitor team sentiment regarding current configuration;
Analyze output quality trends; Implement interpretability assessment
protocols & Identifies transition needs before they become urgent;
surfaces emerging challenges; tracks quality indicators; supports
proactive adjustment \\
\end{longtable}

\paragraph{A.1.4.2 Configuration Flexibility
Strategies}\label{a.1.4.2-configuration-flexibility-strategies}

To enable smooth and deliberate transitions, build flexibility into the
partnership's operating model:

\begin{longtable}[]{@{}
  >{\raggedright\arraybackslash}p{(\columnwidth - 4\tabcolsep) * \real{0.1786}}
  >{\raggedright\arraybackslash}p{(\columnwidth - 4\tabcolsep) * \real{0.4821}}
  >{\raggedright\arraybackslash}p{(\columnwidth - 4\tabcolsep) * \real{0.3393}}@{}}
\toprule\noalign{}
\begin{minipage}[b]{\linewidth}\raggedright
Strategy
\end{minipage} & \begin{minipage}[b]{\linewidth}\raggedright
Implementation Approaches
\end{minipage} & \begin{minipage}[b]{\linewidth}\raggedright
Expected Benefits
\end{minipage} \\
\midrule\noalign{}
\endhead
\bottomrule\noalign{}
\endlastfoot
\textbf{Modal Frameworks} & Clearly delineate configuration
characteristics; Document transition procedures between modes; Develop
configuration-specific protocols; Train teams for multi-modal operation
& Creates clear understanding of different modes; enables intentional
switching; builds operational flexibility; develops team adaptability \\
\textbf{Hybrid Configurations} & Implement domain-specific agency
allocation; Create phase-based configuration variation; Develop
task-specific configuration selection; Design researcher-specific
configuration adaptations & Enables nuanced application of different
configurations; supports context-sensitivity; allows targeted agency
allocation; maximizes configuration benefits \\
\textbf{Configuration Documentation} & Implement configuration decision
logging; Track outcomes across different configurations; Conduct
comparative performance assessment; Document configuration evolution
over time & Creates organizational memory regarding configurations;
enables comparison of effectiveness; supports learning; informs future
decisions \\
\end{longtable}

\paragraph{A.1.4.3 Transition Protocol
Strategies}\label{a.1.4.3-transition-protocol-strategies}

To manage configuration changes effectively and minimize disruption,
establish clear protocols for transitions:

\begin{longtable}[]{@{}
  >{\raggedright\arraybackslash}p{(\columnwidth - 4\tabcolsep) * \real{0.1786}}
  >{\raggedright\arraybackslash}p{(\columnwidth - 4\tabcolsep) * \real{0.4821}}
  >{\raggedright\arraybackslash}p{(\columnwidth - 4\tabcolsep) * \real{0.3393}}@{}}
\toprule\noalign{}
\begin{minipage}[b]{\linewidth}\raggedright
Strategy
\end{minipage} & \begin{minipage}[b]{\linewidth}\raggedright
Implementation Approaches
\end{minipage} & \begin{minipage}[b]{\linewidth}\raggedright
Expected Benefits
\end{minipage} \\
\midrule\noalign{}
\endhead
\bottomrule\noalign{}
\endlastfoot
\textbf{Staged Transition Approaches} & Begin with preparatory
assessment phase; Conduct controlled transition experiments; Implement
limited-scope initial transitions; Progressively expand with ongoing
evaluation & Reduces transition risks; builds experience gradually;
enables adjustment; maintains research momentum \\
\textbf{Reversibility Mechanisms} & Document baselines before
transitions; Build parallel operation capabilities; Establish rollback
protocols; Define threshold-based reversion triggers & Creates safety
nets for transitions; enables swift correction if needed; reduces
transition risk; supports experimentation \\
\textbf{Team Adaptation Support} & Provide role evolution guidance;
Create skill development opportunities; Implement expectation management
approaches; Develop configuration-specific training & Supports human
adaptation to new configurations; reduces transition stress; builds
necessary capabilities; aligns expectations \\
\end{longtable}

\begin{center}\rule{0.5\linewidth}{0.5pt}\end{center}

\begin{quote}
\textbf{REFLECTIVE QUESTIONS: TRANSITION MANAGEMENT}

Consider these questions to strengthen your approach to transition
management:

\begin{enumerate}
\def\labelenumi{\arabic{enumi}.}
\item
  Reviewing the indicators in A.1.4.1, what signals might specifically
  indicate readiness for increased AI agency (Directed → Contributory or
  Contributory → Partnership) in your research context?
\item
  Conversely, what circumstances or indicators (A.1.4.1) might suggest
  benefits from reinforcing human oversight (Partnership → Contributory
  or Contributory → Directed)?
\item
  How can transition protocols (A.1.4.3) be designed to maintain
  research momentum while effectively adapting agency configurations?
\item
  What organizational or team capabilities (e.g., skills, mindset,
  resources related to A.1.4.2) currently enable or hinder smoother
  transitions between configurations?
\item
  How might your team develop organizational memory regarding successful
  and challenging transitions to inform future configuration management?
\end{enumerate}
\end{quote}

\subsection{A.2 Epistemic Dimension Implementation}\label{a.2-epistemic-dimension-implementation}

\begin{quote}
\textbf{AT A GLANCE: EPISTEMIC DIMENSIONS}

The framework identifies six key dimensions of epistemic capability that
emerge in human-AI partnerships:

\textbf{Discovery Axis}

\begin{itemize}
\tightlist
\item
  \textbf{Divergent Intelligence}: Generating novel possibilities
\item
  \textbf{Interpretive Intelligence}: Making sense of complex patterns
  and outputs
\end{itemize}

\textbf{Integration Axis}

\begin{itemize}
\tightlist
\item
  \textbf{Connective Intelligence}: Identifying relationships across
  domains
\item
  \textbf{Synthesis Intelligence}: Building coherent frameworks
\end{itemize}

\textbf{Projection Axis}

\begin{itemize}
\tightlist
\item
  \textbf{Anticipatory Intelligence}: Exploring future trajectories
\item
  \textbf{Axiological Intelligence}: Evaluating significance and value
\end{itemize}

Effective implementation involves deliberately cultivating these
dimensions to develop balanced capability signatures.
\end{quote}

\subsubsection{A.2.1 Discovery Axis: Detailed
Strategies}\label{a.2.1-discovery-axis-detailed-strategies}

\begin{quote}
\textbf{AT A GLANCE: DISCOVERY AXIS}\\
\emph{Cultivating capabilities for generating novel possibilities and
making them interpretable}

\textbf{Key Components:}

-\textbf{Divergent Intelligence}: Techniques for expanding the space of
considered possibilities - \textbf{Interpretive Intelligence}:
Approaches for making complex patterns and outputs understandable
\end{quote}

\paragraph{A.2.1.1 Divergent Intelligence
Strategies}\label{a.2.1.1-divergent-intelligence-strategies}

To enhance the partnership's ability to generate novel possibilities
beyond conventional thinking, implement strategies that systematically
expand the solution space:

\begin{longtable}[]{@{}
  >{\raggedright\arraybackslash}p{(\columnwidth - 4\tabcolsep) * \real{0.1786}}
  >{\raggedright\arraybackslash}p{(\columnwidth - 4\tabcolsep) * \real{0.4821}}
  >{\raggedright\arraybackslash}p{(\columnwidth - 4\tabcolsep) * \real{0.3393}}@{}}
\toprule\noalign{}
\begin{minipage}[b]{\linewidth}\raggedright
Strategy
\end{minipage} & \begin{minipage}[b]{\linewidth}\raggedright
Implementation Approaches
\end{minipage} & \begin{minipage}[b]{\linewidth}\raggedright
Expected Benefits
\end{minipage} \\
\midrule\noalign{}
\endhead
\bottomrule\noalign{}
\endlastfoot
\textbf{Variation Techniques} & Implement parameter variation protocols;
Create constraint relaxation methods; Develop seed diversity approaches;
Apply multi-model comparison techniques & Increases diversity of
outputs; explores solution space systematically; reduces fixation on
familiar patterns; discovers unexpected possibilities \\
\textbf{Possibility Expansion} & Conduct counterfactual exploration
exercises; Apply assumption reversal techniques; Facilitate disciplinary
cross-pollination; Experiment with constraint elimination & Challenges
implicit assumptions; generates radical alternatives; introduces
cross-domain insights; expands conceptual boundaries \\
\textbf{Novelty Assessment} & Conduct literature-based novelty
measurement; Incorporate disciplinary expert assessment; Perform
computational distinctiveness analysis; Evaluate against historical
precedents & Objectively evaluates innovation; clarifies contribution
value; distinguishes true novelty from variation; contextualizes
discoveries \\
\end{longtable}

\paragraph{A.2.1.2 Interpretive Intelligence
Strategies}\label{a.2.1.2-interpretive-intelligence-strategies}

To make sense of complex AI-generated patterns and outputs, develop
approaches that enhance understanding without oversimplifying:

\begin{longtable}[]{@{}
  >{\raggedright\arraybackslash}p{(\columnwidth - 4\tabcolsep) * \real{0.1786}}
  >{\raggedright\arraybackslash}p{(\columnwidth - 4\tabcolsep) * \real{0.4821}}
  >{\raggedright\arraybackslash}p{(\columnwidth - 4\tabcolsep) * \real{0.3393}}@{}}
\toprule\noalign{}
\begin{minipage}[b]{\linewidth}\raggedright
Strategy
\end{minipage} & \begin{minipage}[b]{\linewidth}\raggedright
Implementation Approaches
\end{minipage} & \begin{minipage}[b]{\linewidth}\raggedright
Expected Benefits
\end{minipage} \\
\midrule\noalign{}
\endhead
\bottomrule\noalign{}
\endlastfoot
\textbf{Explanation Protocols} & Generate counterfactual explanations;
Develop visualization approaches; Create narrative explanations;
Implement mechanistic decomposition & Increases understanding of causal
relationships; makes patterns perceptible; connects to human intuition;
breaks complexity into components \\
\textbf{Progressive Disclosure} & Create high-level summaries; Develop
intermediate conceptual explanations; Provide detailed technical
descriptions; Offer complete mechanistic accounts & Addresses different
depths of understanding needs; prevents cognitive overload; enables
appropriate detail level; supports both quick scanning and deep
analysis \\
\textbf{Context-Sensitive Explanations} & Apply discipline-specific
vocabularies; Align with theoretical frameworks; Connect to
methodological traditions; Resonate with epistemic values & Increases
relevance to specific audiences; integrates with existing knowledge;
reduces translation barriers; respects disciplinary norms \\
\end{longtable}

\subsubsection{A.2.2 Integration Axis: Detailed
Strategies}\label{a.2.2-integration-axis-detailed-strategies}

\begin{quote}
\textbf{AT A GLANCE: INTEGRATION AXIS}\\
\emph{Developing capabilities to connect disparate knowledge and
synthesize coherent frameworks}

\textbf{Key Components:}

\begin{itemize}
\tightlist
\item
  \textbf{Connective Intelligence}: Approaches for identifying
  relationships across domains
\item
  \textbf{Synthesis Intelligence}: Methods for building integrated
  explanatory frameworks
\end{itemize}
\end{quote}

\paragraph{A.2.2.1 Connective Intelligence
Strategies}\label{a.2.2.1-connective-intelligence-strategies}

To improve the partnership's capacity to identify meaningful
relationships across domains and knowledge types, implement strategies
that facilitate cross-domain exploration:

\begin{longtable}[]{@{}
  >{\raggedright\arraybackslash}p{(\columnwidth - 4\tabcolsep) * \real{0.1786}}
  >{\raggedright\arraybackslash}p{(\columnwidth - 4\tabcolsep) * \real{0.4821}}
  >{\raggedright\arraybackslash}p{(\columnwidth - 4\tabcolsep) * \real{0.3393}}@{}}
\toprule\noalign{}
\begin{minipage}[b]{\linewidth}\raggedright
Strategy
\end{minipage} & \begin{minipage}[b]{\linewidth}\raggedright
Implementation Approaches
\end{minipage} & \begin{minipage}[b]{\linewidth}\raggedright
Expected Benefits
\end{minipage} \\
\midrule\noalign{}
\endhead
\bottomrule\noalign{}
\endlastfoot
\textbf{Cross-Domain Exposure} & Implement literature scanning
protocols; Provide interdisciplinary database access; Create
cross-disciplinary collaboration structures; Expose partnerships to
methodological diversity & Broadens knowledge horizon; surfaces
unexpected parallels; reduces siloed thinking; introduces diverse
approaches \\
\textbf{Relationship Visualization} & Apply network visualization
techniques; Create conceptual mapping approaches; Develop semantic
relationship diagrams; Implement cross-disciplinary matrices & Makes
abstract relationships perceptible; reveals hidden patterns; supports
spatial understanding; shows multi-dimensional connections \\
\textbf{Connection Validation} & Implement theoretical validation
approaches; Apply empirical testing methods; Conduct expert review
procedures; Assess practical utility & Distinguishes meaningful from
spurious connections; grounds relationships in evidence; validates
through multiple perspectives; ensures practical relevance \\
\end{longtable}

\paragraph{A.2.2.2 Synthesis Intelligence
Strategies}\label{a.2.2.2-synthesis-intelligence-strategies}

To develop the partnership's ability to integrate diverse knowledge into
coherent frameworks, implement approaches that facilitate meaningful
synthesis:

\begin{longtable}[]{@{}
  >{\raggedright\arraybackslash}p{(\columnwidth - 4\tabcolsep) * \real{0.1786}}
  >{\raggedright\arraybackslash}p{(\columnwidth - 4\tabcolsep) * \real{0.4821}}
  >{\raggedright\arraybackslash}p{(\columnwidth - 4\tabcolsep) * \real{0.3393}}@{}}
\toprule\noalign{}
\begin{minipage}[b]{\linewidth}\raggedright
Strategy
\end{minipage} & \begin{minipage}[b]{\linewidth}\raggedright
Implementation Approaches
\end{minipage} & \begin{minipage}[b]{\linewidth}\raggedright
Expected Benefits
\end{minipage} \\
\midrule\noalign{}
\endhead
\bottomrule\noalign{}
\endlastfoot
\textbf{Integration Workshops} & Conduct multi-perspective synthesis
sessions; Compare competing frameworks systematically; Develop
integrative models collaboratively; Validate synthesis through diverse
lenses & Creates opportunities for deliberate integration; evaluates
alternative frameworks; builds collaborative understanding; ensures
robust synthesis \\
\textbf{Multi-Level Modeling} & Develop micro-mechanism models; Create
mid-level theoretical integration; Articulate macro-level principles;
Map connections across levels & Addresses different scales of
understanding; ensures coherence across levels; facilitates appropriate
abstraction; integrates detailed and broad perspectives \\
\textbf{Framework Evaluation} & Measure explanatory power
systematically; Assess parsimony and elegance; Evaluate practical
utility; Generate novel predictions & Ensures framework quality;
balances comprehensiveness with simplicity; connects to practical
application; tests predictive power \\
\end{longtable}

\subsubsection{A.2.3 Projection Axis: Detailed
Strategies}\label{a.2.3-projection-axis-detailed-strategies}

\begin{quote}
\textbf{AT A GLANCE: PROJECTION AXIS}\\
\emph{Building capabilities to explore future possibilities and evaluate
their significance}

\textbf{Key Components:}

\begin{itemize}
\tightlist
\item
  \textbf{Anticipatory Intelligence}: Methods for generating and
  assessing alternative futures
\item
  \textbf{Axiological Intelligence}: Approaches for evaluating
  significance and implications
\end{itemize}
\end{quote}

\paragraph{A.2.3.1 Anticipatory Intelligence
Strategies}\label{a.2.3.1-anticipatory-intelligence-strategies}

To strengthen the partnership's ability to anticipate future
developments and prepare for alternative possibilities, implement
forward-looking approaches:

\begin{longtable}[]{@{}
  >{\raggedright\arraybackslash}p{(\columnwidth - 4\tabcolsep) * \real{0.1786}}
  >{\raggedright\arraybackslash}p{(\columnwidth - 4\tabcolsep) * \real{0.4821}}
  >{\raggedright\arraybackslash}p{(\columnwidth - 4\tabcolsep) * \real{0.3393}}@{}}
\toprule\noalign{}
\begin{minipage}[b]{\linewidth}\raggedright
Strategy
\end{minipage} & \begin{minipage}[b]{\linewidth}\raggedright
Implementation Approaches
\end{minipage} & \begin{minipage}[b]{\linewidth}\raggedright
Expected Benefits
\end{minipage} \\
\midrule\noalign{}
\endhead
\bottomrule\noalign{}
\endlastfoot
\textbf{Scenario Development} & Conduct structured scenario
construction; Create alternative future maps; Model probability
distributions; Develop branching possibility trees & Systematically
explores futures; reduces surprise; provides decision frameworks;
reveals potential consequences \\
\textbf{Temporal Mapping} & Create timeline visualization techniques;
Model evolutionary trajectories; Map decision points systematically;
Identify critical junctures & Clarifies development pathways; reveals
temporal relationships; highlights intervention points; shows
evolutionary trajectories \\
\textbf{Early Indicator Development} & Implement weak signal detection
protocols; Identify leading indicators; Apply trend analysis techniques;
Develop pattern recognition approaches & Enables earlier awareness of
emerging futures; connects present signals to future states; provides
monitoring frameworks; supports proactive response \\
\end{longtable}

\paragraph{A.2.3.2 Axiological Intelligence
Strategies}\label{a.2.3.2-axiological-intelligence-strategies}

To develop the partnership's capacity to evaluate the significance and
implications of potential developments, implement approaches that engage
with deeper values:

\begin{longtable}[]{@{}
  >{\raggedright\arraybackslash}p{(\columnwidth - 4\tabcolsep) * \real{0.1786}}
  >{\raggedright\arraybackslash}p{(\columnwidth - 4\tabcolsep) * \real{0.4821}}
  >{\raggedright\arraybackslash}p{(\columnwidth - 4\tabcolsep) * \real{0.3393}}@{}}
\toprule\noalign{}
\begin{minipage}[b]{\linewidth}\raggedright
Strategy
\end{minipage} & \begin{minipage}[b]{\linewidth}\raggedright
Implementation Approaches
\end{minipage} & \begin{minipage}[b]{\linewidth}\raggedright
Expected Benefits
\end{minipage} \\
\midrule\noalign{}
\endhead
\bottomrule\noalign{}
\endlastfoot
\textbf{Value Dialogue Techniques} & Conduct structured value
elicitation; Implement value prioritization exercises; Apply value
conflict resolution methods; Track value evolution over time & Surfaces
implicit values; clarifies priorities; resolves tensions between values;
recognizes evolving perspectives \\
\textbf{Evaluative Plurality} & Develop multi-criteria assessment
protocols; Implement value perspective rotation; Document competing
value frameworks; Create value synthesis approaches & Prevents narrow
evaluation; honors diverse perspectives; maintains ethical complexity;
supports nuanced assessment \\
\textbf{Stakeholder Inclusion} & Conduct structured stakeholder
consultation; Create value perspective maps; Develop inclusive
evaluation frameworks; Facilitate cross-stakeholder synthesis & Broadens
evaluative perspectives; identifies overlooked concerns; increases
relevance to diverse groups; builds shared understanding \\
\end{longtable}

\subsubsection{A.2.4 Dimensional Assessment: Capability Signature
Mapping}\label{a.2.4-dimensional-assessment-capability-signature-mapping}

\begin{quote}
\textbf{AT A GLANCE: DIMENSIONAL ASSESSMENT}\\
\emph{Developing approaches to map, evaluate, and strategically develop
capabilities across dimensions}

\textbf{Key Components:}

\begin{itemize}
\tightlist
\item
  \textbf{Assessment Strategies}: Methods for evaluating dimensional
  development
\item
  \textbf{Capability Development}: Approaches for strengthening specific
  dimensions
\item
  \textbf{Balance Cultivation}: Techniques for developing synergistic
  capabilities
\end{itemize}
\end{quote}

\paragraph{A.2.4.1 Assessment
Strategies}\label{a.2.4.1-assessment-strategies}

Understanding and visualizing the partnership's strengths and weaknesses
across the epistemic dimensions is key to targeted development:

\begin{longtable}[]{@{}
  >{\raggedright\arraybackslash}p{(\columnwidth - 4\tabcolsep) * \real{0.1786}}
  >{\raggedright\arraybackslash}p{(\columnwidth - 4\tabcolsep) * \real{0.4821}}
  >{\raggedright\arraybackslash}p{(\columnwidth - 4\tabcolsep) * \real{0.3393}}@{}}
\toprule\noalign{}
\begin{minipage}[b]{\linewidth}\raggedright
Strategy
\end{minipage} & \begin{minipage}[b]{\linewidth}\raggedright
Implementation Approaches
\end{minipage} & \begin{minipage}[b]{\linewidth}\raggedright
Expected Benefits
\end{minipage} \\
\midrule\noalign{}
\endhead
\bottomrule\noalign{}
\endlastfoot
\textbf{Dimension Evaluation Protocols} & Create qualitative assessment
frameworks; Develop quantitative measurement approaches; Implement
comparative benchmarking; Track developmental trajectories & Provides
structured evaluation; creates comparable metrics; establishes relative
standing; shows progress over time \\
\textbf{Capability Signature Visualization} & Implement radar chart
visualization; Create temporal evolution maps; Conduct comparative
signature analysis; Identify developmental gaps & Makes abstract
capabilities concrete; shows balance across dimensions; highlights
comparative strengths; reveals improvement opportunities \\
\textbf{Dimensional Balance Assessment} & Systematically identify
capability imbalances; Develop compensatory development plans; Map
synergistic relationships; Assess integration potential & Reveals
unhealthy imbalances; guides development priorities; leverages
relationships between dimensions; supports holistic growth \\
\end{longtable}

\paragraph{A.2.4.2 Capability Development
Planning}\label{a.2.4.2-capability-development-planning}

To systematically strengthen capabilities across dimensions, implement
targeted development approaches:

\begin{longtable}[]{@{}
  >{\raggedright\arraybackslash}p{(\columnwidth - 4\tabcolsep) * \real{0.1786}}
  >{\raggedright\arraybackslash}p{(\columnwidth - 4\tabcolsep) * \real{0.4821}}
  >{\raggedright\arraybackslash}p{(\columnwidth - 4\tabcolsep) * \real{0.3393}}@{}}
\toprule\noalign{}
\begin{minipage}[b]{\linewidth}\raggedright
Strategy
\end{minipage} & \begin{minipage}[b]{\linewidth}\raggedright
Implementation Approaches
\end{minipage} & \begin{minipage}[b]{\linewidth}\raggedright
Expected Benefits
\end{minipage} \\
\midrule\noalign{}
\endhead
\bottomrule\noalign{}
\endlastfoot
\textbf{Targeted Development Strategies} & Create dimension-specific
practice regimens; Implement developmental exercise protocols; Conduct
capability enhancement workshops; Design progressive challenge sequences
& Provides focused capability building; addresses specific weaknesses;
creates systematic improvement; builds progressive mastery \\
\textbf{Integration Planning} & Design cross-dimensional practice
activities; Develop integrative challenge scenarios; Create holistic
capability development contexts; Implement balanced development
sequencing & Builds connections between dimensions; addresses
capabilities holistically; creates realistic development scenarios;
ensures balanced growth \\
\textbf{Development Tracking} & Implement progress visualization
techniques; Track developmental milestones; Conduct comparative
trajectory analysis; Create adaptive planning protocols & Makes progress
visible; provides achievement markers; enables comparative evaluation;
supports flexible development \\
\end{longtable}

\begin{center}\rule{0.5\linewidth}{0.5pt}\end{center}

\begin{quote}
\textbf{REFLECTIVE QUESTIONS: EPISTEMIC DIMENSIONS}

Consider these questions to strengthen your approach to dimensional
development:

\begin{enumerate}
\def\labelenumi{\arabic{enumi}.}
\item
  Which dimensions currently show the greatest development in your
  partnership, and which might benefit from targeted enhancement?
\item
  What developmental imbalances exist across dimensions in your context,
  and how might they affect research quality and outcomes?
\item
  How do your current dimensional strengths and weaknesses align with
  your specific research requirements and goals?
\item
  Which specific approaches from sections A.2.1-A.2.3 might best
  strengthen your underdeveloped dimensions?
\item
  How might you visualize and communicate your partnership's capability
  signature to team members in ways that motivate balanced development?
\end{enumerate}
\end{quote}

\subsection{A.3 Partnership Dynamics Implementation}\label{a.3-partnership-dynamics-implementation}

\begin{quote}
\textbf{AT A GLANCE: PARTNERSHIP DYNAMICS}

The framework addresses three categories of dynamics that shape
partnership evolution:

\textbf{Generative Dynamics}

\begin{itemize}
\tightlist
\item
  \textbf{Transformative Potential}: Capacity for fundamental paradigm
  shifts
\item
  \textbf{Recursive Evolution}: Development through reinforcing feedback
  loops
\end{itemize}

\textbf{Balancing Dynamics}

\begin{itemize}
\tightlist
\item
  \textbf{Temporal Integration}: Harmonizing different temporal frames
\item
  \textbf{Epistemic Ambidexterity}: Balancing innovation and
  interpretation
\end{itemize}

\textbf{Risk Dynamics}

\begin{itemize}
\tightlist
\item
  \textbf{Epistemic Alienation}: Disconnection from meaningful
  understanding
\item
  \textbf{Epistemic Closure}: Narrowing of conceptual possibilities
\end{itemize}

Effective implementation involves cultivating generative forces,
managing tensions, and mitigating risks.
\end{quote}

\subsubsection{A.3.1 Generative Dynamics: Detailed
Strategies}\label{a.3.1-generative-dynamics-detailed-strategies}

\begin{quote}
\textbf{AT A GLANCE: GENERATIVE DYNAMICS}\\
\emph{Cultivating forces that drive innovation, growth, and fundamental
transformation}

\textbf{Key Components:}

\begin{itemize}
\tightlist
\item
  \textbf{Transformative Potential}: Approaches for challenging
  disciplinary paradigms
\item
  \textbf{Recursive Evolution}: Methods for fostering self-reinforcing
  development
\end{itemize}
\end{quote}

\paragraph{A.3.1.1 Transformative Potential
Strategies}\label{a.3.1.1-transformative-potential-strategies}

To foster conditions for breakthrough innovation and sustained
partnership growth, implement approaches that challenge conventional
frameworks:

\begin{longtable}[]{@{}
  >{\raggedright\arraybackslash}p{(\columnwidth - 4\tabcolsep) * \real{0.1786}}
  >{\raggedright\arraybackslash}p{(\columnwidth - 4\tabcolsep) * \real{0.4821}}
  >{\raggedright\arraybackslash}p{(\columnwidth - 4\tabcolsep) * \real{0.3393}}@{}}
\toprule\noalign{}
\begin{minipage}[b]{\linewidth}\raggedright
Strategy
\end{minipage} & \begin{minipage}[b]{\linewidth}\raggedright
Implementation Approaches
\end{minipage} & \begin{minipage}[b]{\linewidth}\raggedright
Expected Benefits
\end{minipage} \\
\midrule\noalign{}
\endhead
\bottomrule\noalign{}
\endlastfoot
\textbf{Paradigm Exploration} & Conduct assumption identification
exercises; Implement paradigmatic boundary testing; Explore alternative
paradigm frameworks; Develop integrative paradigms across domains &
Reveals implicit constraints; challenges foundational assumptions; opens
new conceptual spaces; fosters transdisciplinary innovation \\
\textbf{Constraint Relaxation} & Develop constraint identification
protocols; Implement structured relaxation exercises; Create
unconstrained exploration sessions; Establish reintegration approaches &
Temporarily frees thinking from limitations; allows radical
possibilities; generates unconventional perspectives; supports return to
productive constraints \\
\textbf{Transformation Documentation} & Create conceptual shift mapping;
Document paradigmatic transitions; Visualize epistemic evolution;
Identify transformative moments & Captures fundamental changes; tracks
conceptual evolution; recognizes pivotal insights; supports reflection
on transformation processes \\
\end{longtable}

\paragraph{A.3.1.2 Recursive Evolution
Strategies}\label{a.3.1.2-recursive-evolution-strategies}

To develop partnership capabilities through reinforcing feedback loops,
implement approaches that accelerate co-evolution:

\begin{longtable}[]{@{}
  >{\raggedright\arraybackslash}p{(\columnwidth - 4\tabcolsep) * \real{0.1786}}
  >{\raggedright\arraybackslash}p{(\columnwidth - 4\tabcolsep) * \real{0.4821}}
  >{\raggedright\arraybackslash}p{(\columnwidth - 4\tabcolsep) * \real{0.3393}}@{}}
\toprule\noalign{}
\begin{minipage}[b]{\linewidth}\raggedright
Strategy
\end{minipage} & \begin{minipage}[b]{\linewidth}\raggedright
Implementation Approaches
\end{minipage} & \begin{minipage}[b]{\linewidth}\raggedright
Expected Benefits
\end{minipage} \\
\midrule\noalign{}
\endhead
\bottomrule\noalign{}
\endlastfoot
\textbf{Evolutionary Documentation} & Track interaction patterns
systematically; Document capability development trajectories; Map
relationship evolution visually; Identify developmental phases & Creates
awareness of evolution; reveals growth patterns; supports developmental
understanding; enables strategic adaptation \\
\textbf{Meta-Learning Approaches} & Analyze learning patterns across
interactions; Identify adaptive strategy development; Recognize learning
obstacles; Enhance meta-cognitive awareness & Improves learning about
learning; accelerates adaptation; overcomes developmental barriers;
builds reflective capacity \\
\textbf{Feedback Amplification} & Identify productive feedback patterns;
Design amplification mechanisms; Implement feedback acceleration
techniques; Cultivate virtuoso development cycles & Strengthens positive
dynamics; accelerates growth in capabilities; focuses resources on
productive patterns; creates developmental momentum \\
\end{longtable}

\subsubsection{A.3.2 Balancing Dynamics: Detailed
Strategies}\label{a.3.2-balancing-dynamics-detailed-strategies}

\begin{quote}
\textbf{AT A GLANCE: BALANCING DYNAMICS}\\
\emph{Managing tensions between competing needs to maintain productive
equilibrium}

\textbf{Key Components:}

-\textbf{Temporal Integration}: Methods for working across different
time scales - \textbf{Epistemic Ambidexterity}: Approaches for balancing
innovation and interpretation
\end{quote}

\paragraph{A.3.2.1 Temporal Integration
Strategies}\label{a.3.2.1-temporal-integration-strategies}

To ensure sustainable progress by harmonizing different temporal frames,
implement approaches that integrate multiple timescales:

\begin{longtable}[]{@{}
  >{\raggedright\arraybackslash}p{(\columnwidth - 4\tabcolsep) * \real{0.1786}}
  >{\raggedright\arraybackslash}p{(\columnwidth - 4\tabcolsep) * \real{0.4821}}
  >{\raggedright\arraybackslash}p{(\columnwidth - 4\tabcolsep) * \real{0.3393}}@{}}
\toprule\noalign{}
\begin{minipage}[b]{\linewidth}\raggedright
Strategy
\end{minipage} & \begin{minipage}[b]{\linewidth}\raggedright
Implementation Approaches
\end{minipage} & \begin{minipage}[b]{\linewidth}\raggedright
Expected Benefits
\end{minipage} \\
\midrule\noalign{}
\endhead
\bottomrule\noalign{}
\endlastfoot
\textbf{Temporal Awareness} & Implement temporal horizon expansion
exercises; Conduct multi-timescale analysis; Identify temporal
assumptions; Develop time-scale integration methods & Creates
consciousness of temporal dimensions; reveals assumptions about time;
connects immediate and long-term; supports temporal perspective \\
\textbf{Tempo Modulation} & Establish deliberate deceleration protocols;
Implement accelerated exploration techniques; Create rhythmic
alternation approaches; Develop context-sensitive pacing strategies &
Matches pace to epistemic needs; creates space for reflection; supports
rapid exploration when appropriate; develops rhythmic sensitivity \\
\textbf{Temporal Portfolio} & Design parallel timeframe projects;
Implement temporal diversity planning; Create complementary rhythm
structures; Develop integrated temporal orchestration & Balances
immediate and long-term work; creates temporal resilience; provides
complementary perspectives; enables coordinated progress \\
\end{longtable}

\paragraph{A.3.2.2 Epistemic Ambidexterity
Strategies}\label{a.3.2.2-epistemic-ambidexterity-strategies}

To balance innovation with interpretation, implement approaches that
support both exploration and explanation:

\begin{longtable}[]{@{}
  >{\raggedright\arraybackslash}p{(\columnwidth - 4\tabcolsep) * \real{0.1786}}
  >{\raggedright\arraybackslash}p{(\columnwidth - 4\tabcolsep) * \real{0.4821}}
  >{\raggedright\arraybackslash}p{(\columnwidth - 4\tabcolsep) * \real{0.3393}}@{}}
\toprule\noalign{}
\begin{minipage}[b]{\linewidth}\raggedright
Strategy
\end{minipage} & \begin{minipage}[b]{\linewidth}\raggedright
Implementation Approaches
\end{minipage} & \begin{minipage}[b]{\linewidth}\raggedright
Expected Benefits
\end{minipage} \\
\midrule\noalign{}
\endhead
\bottomrule\noalign{}
\endlastfoot
\textbf{Balance Assessment} & Develop balance metric frameworks; Create
equilibrium evaluation tools; Implement imbalance detection methods;
Design dynamic adjustment indicators & Provides awareness of current
balance; signals potential imbalances; supports targeted intervention;
enables continuous adjustment \\
\textbf{Mode Switching} & Establish transition signaling protocols;
Develop mode-specific practice libraries; Implement context-sensitive
switching triggers; Provide fluid transition training & Facilitates
deliberate mode changes; builds capability for different modes; enables
contextual adaptation; reduces mode-switching friction \\
\textbf{Integration Approaches} & Develop innovative interpretation
frameworks; Create interpretive innovation techniques; Implement
bidirectional practice development; Train simultaneous mode operation &
Transcends mode dichotomy; leverages complementarity; builds integrated
capabilities; enables seamless navigation between modes \\
\end{longtable}

\subsubsection{A.3.3 Risk Dynamics: Detailed
Strategies}\label{a.3.3-risk-dynamics-detailed-strategies}

\begin{quote}
\textbf{AT A GLANCE: RISK DYNAMICS}\\
\emph{Proactively addressing epistemic vulnerabilities that threaten
partnership integrity}

\textbf{Key Components:}

\begin{itemize}
\tightlist
\item
  \textbf{Epistemic Alienation}: Methods for maintaining meaningful
  connection to outputs
\item
  \textbf{Epistemic Closure}: Approaches for preventing narrowing of
  possibilities
\end{itemize}
\end{quote}

\paragraph{A.3.3.1 Epistemic Alienation
Strategies}\label{a.3.3.1-epistemic-alienation-strategies}

To protect against disconnection from meaningful understanding of
partnership outputs, implement approaches that build and maintain
interpretive connection:

\begin{longtable}[]{@{}
  >{\raggedright\arraybackslash}p{(\columnwidth - 4\tabcolsep) * \real{0.1786}}
  >{\raggedright\arraybackslash}p{(\columnwidth - 4\tabcolsep) * \real{0.4821}}
  >{\raggedright\arraybackslash}p{(\columnwidth - 4\tabcolsep) * \real{0.3393}}@{}}
\toprule\noalign{}
\begin{minipage}[b]{\linewidth}\raggedright
Strategy
\end{minipage} & \begin{minipage}[b]{\linewidth}\raggedright
Implementation Approaches
\end{minipage} & \begin{minipage}[b]{\linewidth}\raggedright
Expected Benefits
\end{minipage} \\
\midrule\noalign{}
\endhead
\bottomrule\noalign{}
\endlastfoot
\textbf{Alienation Awareness} & Develop early indicator identification
systems; Create alienation assessment tools; Implement team sentiment
monitoring; Measure epistemic distance systematically & Provides early
warning of disconnection; enables targeted intervention; surfaces
subjective experiences; quantifies interpretability challenges \\
\textbf{Interpretive Anchoring} & Build conceptual bridges
systematically; Connect to familiar frameworks; Implement progressive
extension methods; Apply disciplinary grounding approaches & Maintains
meaningful connections; prevents excessive abstraction; builds from
known to unknown; ensures disciplinary relevance \\
\textbf{Ownership Cultivation} & Create interpretive participation
structures; Design co-creation processes; Implement attribution clarity
protocols; Document meaningful influence systematically & Builds
psychological ownership; reduces alienation risk; clarifies human
contributions; creates sense of agency \\
\end{longtable}

\paragraph{A.3.3.2 Epistemic Closure
Strategies}\label{a.3.3.2-epistemic-closure-strategies}

To prevent narrowing of conceptual possibilities and reinforcement of
biases, implement approaches that maintain openness:

\begin{longtable}[]{@{}
  >{\raggedright\arraybackslash}p{(\columnwidth - 4\tabcolsep) * \real{0.1786}}
  >{\raggedright\arraybackslash}p{(\columnwidth - 4\tabcolsep) * \real{0.4821}}
  >{\raggedright\arraybackslash}p{(\columnwidth - 4\tabcolsep) * \real{0.3393}}@{}}
\toprule\noalign{}
\begin{minipage}[b]{\linewidth}\raggedright
Strategy
\end{minipage} & \begin{minipage}[b]{\linewidth}\raggedright
Implementation Approaches
\end{minipage} & \begin{minipage}[b]{\linewidth}\raggedright
Expected Benefits
\end{minipage} \\
\midrule\noalign{}
\endhead
\bottomrule\noalign{}
\endlastfoot
\textbf{Diversity Exposure} & Implement contrasting viewpoint protocols;
Provide alternative framework exposure; Introduce methodological
diversity; Facilitate disciplinary cross-pollination & Prevents echo
chambers; introduces novel perspectives; challenges emerging consensus;
maintains conceptual diversity \\
\textbf{Assumption Surfacing} & Conduct assumption excavation workshops;
Apply presupposition testing approaches; Create background belief
mapping; Facilitate tacit knowledge articulation & Reveals hidden
constraints; enables critical examination; prevents implicit narrowing;
surfaces unconscious limitations \\
\textbf{Counterargument Generation} & Implement structured opposition
protocols; Create devil's advocate frameworks; Require alternative
explanations; Apply disconfirmation strategies & Systematically
challenges emerging consensus; strengthens thinking; prevents premature
closure; maintains critical examination \\
\end{longtable}

\subsubsection{A.3.4 Dynamics Interaction
Management}\label{a.3.4-dynamics-interaction-management}

\begin{quote}
\textbf{AT A GLANCE: DYNAMICS INTERACTION MANAGEMENT}\\
\emph{Understanding and managing complex interactions between different
partnership dynamics}

\textbf{Key Components:}

\begin{itemize}
\tightlist
\item
  \textbf{Dynamics Mapping}: Visualizing relationships and influences
\item
  \textbf{Threshold Monitoring}: Identifying approaching critical
  transitions
\item
  \textbf{Strategic Intervention}: Implementing targeted adjustments
\end{itemize}
\end{quote}

Recognizing that partnership dynamics interact in complex ways,
implement strategies for understanding these interactions and
intervening strategically:

\begin{longtable}[]{@{}
  >{\raggedright\arraybackslash}p{(\columnwidth - 4\tabcolsep) * \real{0.1786}}
  >{\raggedright\arraybackslash}p{(\columnwidth - 4\tabcolsep) * \real{0.4821}}
  >{\raggedright\arraybackslash}p{(\columnwidth - 4\tabcolsep) * \real{0.3393}}@{}}
\toprule\noalign{}
\begin{minipage}[b]{\linewidth}\raggedright
Strategy
\end{minipage} & \begin{minipage}[b]{\linewidth}\raggedright
Implementation Approaches
\end{minipage} & \begin{minipage}[b]{\linewidth}\raggedright
Expected Benefits
\end{minipage} \\
\midrule\noalign{}
\endhead
\bottomrule\noalign{}
\endlastfoot
\textbf{Dynamics Mapping} & Create interaction network visualizations;
Identify reinforcement patterns; Map countervailing forces; Develop
system dynamics models & Reveals complex relationships; shows cascade
effects; identifies potential interventions; builds systemic
understanding \\
\textbf{Threshold Monitoring} & Implement early warning indicator
systems; Measure threshold proximity; Assess transition preparedness;
Identify intervention opportunities & Provides advance notice of
transitions; enables proactive response; ensures readiness for change;
identifies optimal intervention timing \\
\textbf{Strategic Intervention Design} & Identify high-leverage
intervention points; Develop adaptive intervention sequences; Assess
contextual sensitivity factors; Monitor intervention impacts
systematically & Maximizes intervention efficacy; adapts to evolving
conditions; considers systemic context; tracks intervention outcomes \\
\end{longtable}

\begin{center}\rule{0.5\linewidth}{0.5pt}\end{center}

\begin{quote}
\textbf{REFLECTIVE QUESTIONS: PARTNERSHIP DYNAMICS}

Consider these questions to strengthen your approach to managing
partnership dynamics:

\begin{enumerate}
\def\labelenumi{\arabic{enumi}.}
\item
  How do generative dynamics (A.3.1) currently manifest in your
  partnership, and which strategies might further cultivate
  transformative potential?
\item
  What balancing mechanisms (A.3.2) have proven most effective in
  maintaining productive equilibrium in your context?
\item
  Which risk dynamics (A.3.3) require the most active management in your
  partnership, and what specific mitigation strategies might be most
  appropriate?
\item
  What interdependencies exist between different dynamics in your
  partnership, and how might strategic interventions (A.3.4) address
  multiple dynamics simultaneously?
\item
  What indicators or thresholds would suggest a need for significant
  adjustment to your dynamics management approach?
\end{enumerate}
\end{quote}

\subsection{A.4 Implementation Mapping}\label{a.4-implementation-mapping}

\begin{quote}
\textbf{AT A GLANCE: IMPLEMENTATION MAPPING}

These mapping tools help translate conceptual understanding into
practical application by showing:

\begin{itemize}
\tightlist
\item
  How dimensions manifest differently across agency configurations
\item
  How dynamics vary across different configurations
\item
  How to select implementation strategies based on context
\end{itemize}

Use these matrices and decision frameworks to guide initial planning and
adaptation as partnerships evolve.
\end{quote}

\subsubsection{A.4.1 Dimension-Configuration
Matrix}\label{a.4.1-dimension-configuration-matrix}

This matrix illustrates how strategies for cultivating Epistemic
Dimensions might differ depending on the active Agency Configuration.
Use it to identify configuration-specific approaches:

\begin{longtable}[]{@{}
  >{\raggedright\arraybackslash}p{(\columnwidth - 6\tabcolsep) * \real{0.1618}}
  >{\raggedright\arraybackslash}p{(\columnwidth - 6\tabcolsep) * \real{0.2500}}
  >{\raggedright\arraybackslash}p{(\columnwidth - 6\tabcolsep) * \real{0.3088}}
  >{\raggedright\arraybackslash}p{(\columnwidth - 6\tabcolsep) * \real{0.2794}}@{}}
\toprule\noalign{}
\begin{minipage}[b]{\linewidth}\raggedright
Dimension
\end{minipage} & \begin{minipage}[b]{\linewidth}\raggedright
Directed Agency
\end{minipage} & \begin{minipage}[b]{\linewidth}\raggedright
Contributory Agency
\end{minipage} & \begin{minipage}[b]{\linewidth}\raggedright
Partnership Agency
\end{minipage} \\
\midrule\noalign{}
\endhead
\bottomrule\noalign{}
\endlastfoot
\textbf{Divergent Intelligence} & Controlled variation within defined
parameters & Designated exploration spaces for AI-initiated directions &
Co-evolutionary variation techniques \\
\textbf{Interpretive Intelligence} & Standardized explanation protocols
& Bidirectional clarification processes & Emergent interpretive
frameworks \\
\textbf{Connective Intelligence} & Human-directed cross-domain mapping &
AI-suggested unexpected connections & Co-created relationship
identification \\
\textbf{Synthesis Intelligence} & Human framework with AI components &
Hybrid frameworks with distributed authorship & Emergent frameworks
transcending individual contributions \\
\textbf{Anticipatory Intelligence} & Structured future projection &
Complementary human-AI scenario development & Integrated temporal
exploration \\
\textbf{Axiological Intelligence} & Human-defined value application &
Collaborative value assessment & Co-evolutionary value development \\
\end{longtable}

\subsubsection{A.4.2 Dynamics-Configuration
Matrix}\label{a.4.2-dynamics-configuration-matrix}

This matrix highlights how the manifestation and management of
Partnership Dynamics can vary across Agency Configurations. Use it to
anticipate and address configuration-specific dynamic patterns:

\begin{longtable}[]{@{}
  >{\raggedright\arraybackslash}p{(\columnwidth - 6\tabcolsep) * \real{0.1364}}
  >{\raggedright\arraybackslash}p{(\columnwidth - 6\tabcolsep) * \real{0.2576}}
  >{\raggedright\arraybackslash}p{(\columnwidth - 6\tabcolsep) * \real{0.3182}}
  >{\raggedright\arraybackslash}p{(\columnwidth - 6\tabcolsep) * \real{0.2879}}@{}}
\toprule\noalign{}
\begin{minipage}[b]{\linewidth}\raggedright
Dynamic
\end{minipage} & \begin{minipage}[b]{\linewidth}\raggedright
Directed Agency
\end{minipage} & \begin{minipage}[b]{\linewidth}\raggedright
Contributory Agency
\end{minipage} & \begin{minipage}[b]{\linewidth}\raggedright
Partnership Agency
\end{minipage} \\
\midrule\noalign{}
\endhead
\bottomrule\noalign{}
\endlastfoot
\textbf{Transformative Potential} & Limited to efficiency within
paradigms & Potential for paradigm extension & Fundamental paradigm
reconceptualization \\
\textbf{Recursive Evolution} & Linear capability improvement &
Bidirectional adaptation & Complex co-evolutionary development \\
\textbf{Temporal Integration} & Separated timeframes & Coordinated
temporal operations & Integrated multi-temporal functioning \\
\textbf{Epistemic Ambidexterity} & Exploration-explanation separation &
Structured alternation & Simultaneous dual processing \\
\textbf{Epistemic Alienation Risk} & Minimal alienation risk & Moderate
alienation in novel domains & Significant risk requiring active
management \\
\textbf{Epistemic Closure Risk} & Risk of reinforcing existing biases &
Risk of selective incorporation & Risk of novel but narrowing
trajectories \\
\end{longtable}

\subsubsection{A.4.3 Comprehensive Implementation Strategy
Selection}\label{a.4.3-comprehensive-implementation-strategy-selection}

This decision tree offers a starting point for selecting implementation
strategies by considering key contextual factors. Use it to guide
initial planning, recognizing that strategies will need ongoing
adaptation:

\begin{quote}
\textbf{DECISION TREE: STRATEGY SELECTION}

\textbf{1. Research Phase Assessment} (Consider the primary goal of the
current work)

\begin{itemize}
\tightlist
\item
  \textbf{Exploration Phase} → Focus on Divergent Intelligence and
  Transformative Potential
\item
  \textbf{Development Phase} → Focus on Connective Intelligence and
  Synthesis Intelligence
\item
  \textbf{Validation Phase} → Focus on Interpretive Intelligence and
  Epistemic Ambidexterity
\item
  \textbf{Application Phase} → Focus on Anticipatory Intelligence and
  Axiological Intelligence
\end{itemize}

\textbf{2. Team Capability Assessment}

\begin{itemize}
\tightlist
\item
  \textbf{Limited AI Experience} → Begin with Directed Agency strategies
\item
  \textbf{Intermediate AI Experience} → Incorporate Contributory Agency
  strategies
\item
  \textbf{Advanced AI Experience} → Consider Partnership Agency
  strategies
\end{itemize}

\textbf{3. Organizational Context Assessment}

\begin{itemize}
\tightlist
\item
  \textbf{Traditional Academic Environment} → Emphasize Interpretive
  Intelligence and Documentation
\item
  \textbf{Corporate Research Environment} → Emphasize Anticipatory
  Intelligence and Efficiency
\item
  \textbf{Interdisciplinary Environment} → Emphasize Connective
  Intelligence and Integration
\item
  \textbf{Exploratory Research Unit} → Emphasize Divergent Intelligence
  and Transformative Potential
\end{itemize}

\textbf{4. Risk Profile Assessment}

\begin{itemize}
\tightlist
\item
  \textbf{High Accountability Contexts} → Prioritize Epistemic
  Alienation and Closure mitigation
\item
  \textbf{Innovation-Focused Contexts} → Prioritize Generative Dynamics
  cultivation
\item
  \textbf{Resource-Constrained Contexts} → Prioritize Balancing Dynamics
  management
\item
  \textbf{Public-Facing Contexts} → Prioritize Interpretive Intelligence
  development
\end{itemize}
\end{quote}

\subsection{A.5 Reflective Assessment Questions}\label{a.5-reflective-assessment-questions}

\begin{quote}
\textbf{AT A GLANCE: REFLECTIVE ASSESSMENT}

These question sets support ongoing reflection and adaptation across:

\begin{itemize}
\tightlist
\item
  \textbf{Agency Configuration} - Examining how agency is distributed
  and managed
\item
  \textbf{Epistemic Dimensions} - Evaluating capability development
  across dimensions
\item
  \textbf{Partnership Dynamics} - Assessing the evolutionary forces and
  risks shaping the partnership
\end{itemize}

Use these questions periodically to foster team dialogue and guide
continuous improvement.
\end{quote}

\subsubsection{A.5.1 Agency Configuration
Questions}\label{a.5.1-agency-configuration-questions}

\begin{quote}
\textbf{REFLECTIVE QUESTIONS: AGENCY CONFIGURATION}

\begin{enumerate}
\def\labelenumi{\arabic{enumi}.}
\tightlist
\item
  How do current agency configurations align with research phase
  requirements?
\item
  What signals indicate readiness for configuration transitions?
\item
  How effectively do team members navigate between different
  configurations?
\item
  What organizational factors enable or constrain particular
  configurations?
\item
  How might configuration choices influence the quality and nature of
  research outputs?
\item
  What verification approaches are appropriate for each configuration?
\item
  How do different configurations affect team dynamics and roles?
\item
  What documentation approaches best capture configuration evolution?
\item
  How do disciplinary traditions influence appropriate configuration
  choices?
\item
  What ethical considerations arise in different agency configurations?
\end{enumerate}
\end{quote}

\subsubsection{A.5.2 Epistemic Dimension
Questions}\label{a.5.2-epistemic-dimension-questions}

\begin{quote}
\textbf{REFLECTIVE QUESTIONS: EPISTEMIC DIMENSIONS}

\begin{enumerate}
\def\labelenumi{\arabic{enumi}.}
\tightlist
\item
  Which dimensions currently show the greatest development in your
  partnership?
\item
  What developmental imbalances exist across dimensions, and how might
  they affect research?
\item
  How do dimensional strengths and weaknesses align with research
  requirements?
\item
  What approaches might help strengthen underdeveloped dimensions?
\item
  How do dimensional capabilities manifest differently across research
  phases?
\item
  What interdependencies exist between dimensions in your specific
  context?
\item
  How might disciplinary traditions influence dimensional development
  priorities?
\item
  What organizational factors enable or constrain dimensional
  development?
\item
  How do dimensional capabilities influence research trajectory and
  outputs?
\item
  What visualization approaches best capture dimensional development for
  team reflection?
\end{enumerate}
\end{quote}

\subsubsection{A.5.3 Partnership Dynamics
Questions}\label{a.5.3-partnership-dynamics-questions}

\begin{quote}
\textbf{REFLECTIVE QUESTIONS: PARTNERSHIP DYNAMICS}

\begin{enumerate}
\def\labelenumi{\arabic{enumi}.}
\tightlist
\item
  How do generative dynamics currently manifest in your partnership?
\item
  What balancing mechanisms help maintain productive equilibrium?
\item
  What risk dynamics require active management in your context?
\item
  How do organizational structures influence partnership dynamics?
\item
  What threshold effects have you observed in your partnership
  evolution?
\item
  How do different dynamics interact in your specific research context?
\item
  What indicators suggest approaching critical thresholds in partnership
  dynamics?
\item
  How might disciplinary traditions influence which dynamics
  predominate?
\item
  What documentation approaches best capture dynamic evolution?
\item
  How do power structures influence the expression and management of
  dynamics?
\end{enumerate}
\end{quote}

\subsection{A.6 Assessment Protocols}\label{a.6-assessment-protocols}

\begin{quote}
\textbf{AT A GLANCE: ASSESSMENT PROTOCOLS}

These structured assessment approaches provide systematic methods for:

\begin{itemize}
\tightlist
\item
  \textbf{Capability Signature Assessment} - Evaluating partnership
  capabilities across dimensions
\item
  \textbf{Risk Dynamic Assessment} - Identifying and managing potential
  vulnerabilities
\item
  \textbf{Organizational Readiness Assessment} - Determining
  preparedness for implementation
\end{itemize}

Each protocol follows a five-phase process: Preparation, Data
Collection, Analysis, Synthesis, and Implementation.
\end{quote}

\subsubsection{A.6.1 Capability Signature
Assessment}\label{a.6.1-capability-signature-assessment}

The following protocol enables systematic evaluation of partnership
capabilities across dimensions:

\begin{longtable}[]{@{}
  >{\raggedright\arraybackslash}p{(\columnwidth - 4\tabcolsep) * \real{0.2188}}
  >{\raggedright\arraybackslash}p{(\columnwidth - 4\tabcolsep) * \real{0.4688}}
  >{\raggedright\arraybackslash}p{(\columnwidth - 4\tabcolsep) * \real{0.3125}}@{}}
\toprule\noalign{}
\begin{minipage}[b]{\linewidth}\raggedright
Phase
\end{minipage} & \begin{minipage}[b]{\linewidth}\raggedright
Key Activities
\end{minipage} & \begin{minipage}[b]{\linewidth}\raggedright
Outcomes
\end{minipage} \\
\midrule\noalign{}
\endhead
\bottomrule\noalign{}
\endlastfoot
\textbf{1. Preparatory Phase} & Assemble evaluation team with both
direct participants and external observers; Select appropriate
assessment techniques for each dimension; Establish baseline
expectations and contextual considerations; Determine appropriate
assessment intervals and documentation approaches & Well-defined
assessment framework; clear evaluation criteria; established team roles;
appropriate contextual considerations \\
\textbf{2. Data Collection Phase} & Gather evidence through direct
performance observations; Analyze outputs for dimensional capabilities;
Review process documentation; Conduct team member interviews; Collect
external stakeholder perspectives & Comprehensive data across
dimensions; multiple perspective inputs; qualitative and quantitative
evidence; contextual observations \\
\textbf{3. Analysis Phase} & Evaluate dimensional development using
calibrated rubrics; Identify capability strengths, weaknesses, and
imbalances; Examine interdimensional relationships and dependencies;
Assess configuration appropriateness for current phase & Dimensional
capability profile; identified strengths and weaknesses; understanding
of interdimensional relationships; configuration assessment \\
\textbf{4. Synthesis Phase} & Create visual capability signature
representation; Develop integrated capability assessment; Identify
priority areas for development; Design targeted enhancement strategies &
Visual capability map; prioritized development areas; integrated
understanding; actionable enhancement strategies \\
\textbf{5. Implementation Phase} & Communicate assessment findings to
team members; Develop specific capability development plans; Implement
selected enhancement strategies; Establish monitoring approaches for
tracking progress & Shared understanding of capabilities; targeted
development plans; implemented enhancements; progress tracking
mechanisms \\
\end{longtable}

\subsubsection{A.6.2 Risk Dynamic
Assessment}\label{a.6.2-risk-dynamic-assessment}

The following protocol enables systematic evaluation of risk dynamics:

\begin{longtable}[]{@{}
  >{\raggedright\arraybackslash}p{(\columnwidth - 4\tabcolsep) * \real{0.2188}}
  >{\raggedright\arraybackslash}p{(\columnwidth - 4\tabcolsep) * \real{0.4688}}
  >{\raggedright\arraybackslash}p{(\columnwidth - 4\tabcolsep) * \real{0.3125}}@{}}
\toprule\noalign{}
\begin{minipage}[b]{\linewidth}\raggedright
Phase
\end{minipage} & \begin{minipage}[b]{\linewidth}\raggedright
Key Activities
\end{minipage} & \begin{minipage}[b]{\linewidth}\raggedright
Outcomes
\end{minipage} \\
\midrule\noalign{}
\endhead
\bottomrule\noalign{}
\endlastfoot
\textbf{1. Preparatory Phase} & Identify risk dimensions particularly
relevant to context; Select appropriate assessment techniques for each
dimension; Establish risk tolerance thresholds; Determine assessment
scheduling and documentation approaches & Focused risk assessment
framework; appropriate evaluation methods; clear risk thresholds;
scheduled assessment plan \\
\textbf{2. Data Collection Phase} & Conduct team sentiment analysis;
Assess output diversity patterns; Evaluate interpretive alignment; Map
dependency relationships; Examine attribution patterns & Comprehensive
risk indicators; multiple data sources; objective and subjective
measures; contextual risk factors \\
\textbf{3. Analysis Phase} & Evaluate risk levels using calibrated
rubrics; Identify emerging risk patterns and accelerants; Examine
interactions between different risk dynamics; Assess current mitigation
effectiveness & Risk profile across dimensions; identified emerging
patterns; understanding of risk interactions; mitigation effectiveness
evaluation \\
\textbf{4. Synthesis Phase} & Create visual risk profile representation;
Develop integrated risk assessment; Identify priority areas for
mitigation; Design targeted intervention strategies & Visual risk map;
prioritized mitigation areas; integrated understanding; actionable
intervention strategies \\
\textbf{5. Implementation Phase} & Communicate risk assessment findings;
Develop specific risk management plans; Implement selected mitigation
strategies; Establish monitoring approaches for tracking risk evolution
& Shared understanding of risks; targeted mitigation plans; implemented
interventions; risk evolution tracking mechanisms \\
\end{longtable}

\subsubsection{A.6.3 Organizational Readiness
Assessment}\label{a.6.3-organizational-readiness-assessment}

The following protocol enables systematic evaluation of organizational
readiness for implementing the Cognitio Emergens framework:

\begin{longtable}[]{@{}
  >{\raggedright\arraybackslash}p{(\columnwidth - 4\tabcolsep) * \real{0.2188}}
  >{\raggedright\arraybackslash}p{(\columnwidth - 4\tabcolsep) * \real{0.4688}}
  >{\raggedright\arraybackslash}p{(\columnwidth - 4\tabcolsep) * \real{0.3125}}@{}}
\toprule\noalign{}
\begin{minipage}[b]{\linewidth}\raggedright
Phase
\end{minipage} & \begin{minipage}[b]{\linewidth}\raggedright
Key Activities
\end{minipage} & \begin{minipage}[b]{\linewidth}\raggedright
Outcomes
\end{minipage} \\
\midrule\noalign{}
\endhead
\bottomrule\noalign{}
\endlastfoot
\textbf{1. Preparatory Phase} & Identify key organizational dimensions
relevant to implementation; Select appropriate assessment techniques for
each dimension; Establish readiness thresholds and considerations;
Determine assessment scope and documentation approaches & Focused
readiness framework; appropriate evaluation methods; clear readiness
thresholds; well-defined assessment scope \\
\textbf{2. Data Collection Phase} & Conduct leadership interviews;
Analyze relevant policies; Assess resource availability; Evaluate
organizational culture; Examine existing practices & Comprehensive
readiness indicators; multiple data sources; understanding of
organizational context; practice baseline \\
\textbf{3. Analysis Phase} & Evaluate readiness levels across
dimensions; Identify organizational strengths, obstacles, and leverages;
Examine interactions between organizational factors; Assess
implementation feasibility for components & Organizational readiness
profile; identified strengths and barriers; understanding of
organizational dynamics; feasibility assessment \\
\textbf{4. Synthesis Phase} & Create visual organizational readiness
representation; Develop integrated readiness assessment; Identify
priority areas for development; Design staged implementation strategy &
Visual readiness map; prioritized development areas; integrated
understanding; phased implementation approach \\
\textbf{5. Implementation Phase} & Communicate readiness findings to
stakeholders; Develop specific organizational development plans;
Implement preparatory interventions; Establish monitoring for
organizational evolution & Shared understanding of readiness; targeted
development plans; implemented preparations; organizational tracking
mechanisms \\
\end{longtable}

\begin{center}\rule{0.5\linewidth}{0.5pt}\end{center}

These comprehensive implementation strategies, assessment tools, and
reflective questions provide detailed guidance for applying the Cognitio
Emergens framework in diverse research contexts. They are intended as
resources to be selectively applied rather than comprehensive
requirements, supporting thoughtful adaptation to specific disciplinary
traditions, organizational contexts, and research needs.
\end{document}